\documentclass{article}

\usepackage{arxiv}

\usepackage[utf8]{inputenc}
\usepackage[T1]{fontenc}
\usepackage{microtype}
\usepackage{mathptmx}

\usepackage{amsmath,amssymb,amsfonts}
\usepackage{bm}
\usepackage{nicefrac}

\usepackage{graphicx}
\usepackage{svg}
\usepackage{booktabs}
\usepackage{dcolumn}
\usepackage{float}
\usepackage{subcaption}

\usepackage{tikz}
\usetikzlibrary{
    shapes.geometric,
    arrows.meta,
    positioning,
    calc,
    backgrounds,
    fit
}

\usepackage{natbib}
\usepackage{url}
\usepackage{doi}
\usepackage{hyperref}

\title{
Extreme-Scale Super-Resolution of
Homogeneous Turbulence from a
Single Realization
}

\author{
Arth Sojitra\thanks{
These authors contributed equally to this work.
} \\
Department of Mechanical and Aerospace Engineering\\
University of Tennessee, Knoxville\\
Knoxville, TN 37996, USA\\
\texttt{asojitra@vols.utk.edu}
\And
Muhammad Abid\footnotemark[1] \\
Department of Mechanical and Aerospace Engineering\\
University of Tennessee, Knoxville\\
Knoxville, TN 37996, USA\\
\texttt{mabid@vols.utk.edu}
\And
Omer San\\
Department of Mechanical and Aerospace Engineering\\
University of Tennessee, Knoxville\\
Knoxville, TN 37996, USA\\
\texttt{osan@utk.edu}
}

\date{}

\hypersetup{
    pdftitle={
    Extreme-Scale Super-Resolution of
    Homogeneous Turbulence from a
    Single Realization
    },
    pdfauthor={
    Arth Sojitra, Muhammad Abid, Omer San
    },
    pdfsubject={
    Scientific Machine Learning,
    Turbulence Super-Resolution
    },
    pdfkeywords={
    Turbulence,
    Super-Resolution,
    Neural Operators,
    Scientific Machine Learning,
    Computational Fluid Dynamics
    }
}

\begin{document}

\maketitle


\begin{abstract}
\noindent
Homogeneous turbulent flows exhibit recurring multiscale structures across spatially separated regions, suggesting that a single flow realization may contain sufficient information to learn representations of unresolved scales. We investigate this hypothesis through an extreme super-resolution framework in which a wide-activation deep super-resolution (WDSR) network is trained exclusively on overlapping patches extracted from a single high-resolution turbulent realization. Previously unseen flow fields are reconstructed through patchwise prediction followed by Hann-window-weighted overlap-add assembly. The framework is evaluated for two-dimensional homogeneous isotropic turbulence at super-resolution factors of $\times 32$ and $\times 64$, representing severe spatial under-resolution. Despite the limited training data, the proposed approach reconstructs coherent vortical structures and recovers fine-scale features more accurately than conventional interpolation. Spectral and statistical analyses further demonstrate improved recovery of high-wavenumber energy content and multiscale flow characteristics. To assess the effectiveness of exploiting internal spatial information, we compare the proposed framework against WDSR models trained on $1$, $25$, $50$, and $100$ complete high-resolution snapshots without patch extraction. This comparison distinguishes the benefits of localized learning from those associated with increasing the number of training realizations. The results demonstrate that exploiting spatial redundancy within a single turbulent realization enables accurate reconstruction of previously unseen flow fields, even under extreme spatial under-resolution. These findings establish single-realization patch-based learning as a promising data-efficient approach for turbulent super-resolution when high-resolution training data are scarce.
\end{abstract}


\section{Introduction}\label{sec:introduction}

Turbulence is characterized by a hierarchy of interacting spatial and temporal scales, with nonlinear transfer, coherent structures, and intermittency governing transport and mixing \cite{kolmogorov1941local,kraichnan1967inertial,batchelor1953theory,kolmogorov1995turbulence,pope2000turbulent,davidson2004turbulence}. In two-dimensional turbulence, the simultaneous constraints imposed by energy and enstrophy conservation lead to spectral-transfer mechanisms distinct from their three-dimensional counterparts \cite{kraichnan1967inertial,batchelor1969computation}. Direct numerical simulation (DNS) offers detailed representations of these processes, but its cost grows rapidly with Reynolds number and the number of active degrees of freedom \cite{moin1998direct,kaneda2003energy,ishihara2009study}. Experiments are likewise limited by the spatial and temporal resolution of measurement systems \cite{adrian1991particle,raffel2018particle}. Thus, high-resolution flow fields are costly to obtain, while available observations are frequently coarse, noisy, or incomplete.

Reconstructing unresolved scales from coarse measurements is related to the longstanding problem of scale separation in turbulence modeling. Large-eddy simulation (LES) retains resolved motions while modeling the influence of filtered scales through closure assumptions \cite{smagorinsky1963general,germano1991dynamic,meneveau2000scale,sagaut2006large}. Super-resolution instead seeks an explicit fine-grid field consistent with the coarse observation and additional learned information. Crucially, this inverse problem is underdetermined: many distinct high-resolution fields can produce the same coarse measurement. Consequently, interpolation cannot by itself restore discarded information, and even a learned reconstruction must be understood as an inference conditioned on its training prior rather than an exact inversion of coarse graining.

Machine learning has created new opportunities to learn relationships between resolved and unresolved turbulent motions \cite{duraisamy2019turbulence,brunton2020machine, kutz2017deep, vinuesa2023transformative}. Early investigations demonstrated the effectiveness of convolutional neural networks for super-resolution reconstruction of turbulent flow fields, establishing that learned multiscale representations can outperform conventional interpolation in recovering coherent vortical structures and unresolved spatial features \cite{fukami2019superresolution}. Subsequent developments extended these methodologies to spatio-temporal super-resolution, enabling the reconstruction of turbulent fields from observations with limited spatial and temporal resolution \cite{fukami2021spatiotemporal}. The broader development of machine-learning-based super-resolution methodologies for fluid flows, including their architectures, reconstruction capabilities, and practical limitations, has also been systematically reviewed \cite{fukami2023survey}. More recently, spectrally informed multiresolution neural operators have been investigated for turbulent flow super-resolution~\cite{abid2026simr}. Furthermore, convolutional neural networks have been employed to reconstruct wall-bounded turbulent flows from limited measurements, demonstrating the potential of data-driven approaches for recovering complex flow structures from incomplete observations \cite{guastoni2021convolutional}. Recent investigations have examined the influence of adversarial training on turbulent super-resolution, highlighting its potential to improve the reconstruction of small-scale structures and turbulence statistics beyond conventional pointwise reconstruction objectives~\cite{nista2024influence}. Physically constrained generative adversarial networks have also been investigated for reconstructing large missing regions in turbulent flow fields, demonstrating the potential of incorporating physical constraints into data-driven reconstruction of spatially incomplete turbulent observations~\cite{zheng2024high}.

Generative learning approaches have further expanded the
scope of turbulent flow reconstruction by moving beyond
the exclusive minimization of pointwise reconstruction
errors toward the recovery of statistically consistent
fine-scale structures. In particular, unsupervised
super-resolution frameworks based on cycle-consistent
generative adversarial networks (CycleGANs) have
demonstrated the feasibility of learning reconstruction
mappings from unpaired turbulent flow fields, including
homogeneous isotropic turbulence and wall-bounded flows
\cite{kim2021unsupervised}. Generative adversarial networks
have also been employed to reconstruct turbulent flow
fields across different Reynolds numbers, demonstrating
their potential for recovering unresolved spatial
structures and reproducing turbulence statistics
\cite{yousif2022gan}. Beyond conventional super-resolution,
generative models have been investigated for the
reconstruction of spatially incomplete turbulent fields,
including rotating turbulence, where comparisons
against proper orthogonal decomposition have revealed
important differences between instantaneous
reconstruction accuracy and multiscale statistical
fidelity \cite{li2023multiscale}. Similarly, physically
constrained generative adversarial networks have been
developed to reconstruct large missing regions in
compressible isotropic turbulence, illustrating the
potential of incorporating physical information into
generative flow reconstruction
\cite{tretiak2022physics}. Recent investigations
have further examined the influence of adversarial
training on turbulent super-resolution, demonstrating
that discriminator-based learning can improve the
reconstruction of small-scale gradients and
intermittency-related statistics, although
generalization to unseen flow conditions remains
challenging \cite{nista2024influence}. Physics-informed
three-dimensional generative adversarial networks have
also been explored for the reconstruction of
turbulent channel flows from limited-resolution
observations \cite{ward2023physicsinformed}.

Physics-informed methods provide a complementary route to constraining the reconstruction problem. Physics-informed neural networks incorporate differential-equation information into learning objectives \cite{raissi2019physics}; physics-informed convolutional approaches have been developed for super-resolution and denoising of fluid flows \cite{gao2021physics}. Neural operators, including DeepONet and Fourier neural operators, offer alternative formulations for learning mappings between function spaces \cite{lu2021deeponet,li2021fourier,kovachki2023neural}. Recent extensions incorporating Fourier and Chebyshev spectral embeddings, as well as hybrid wavelet-Laplace representations, have further advanced operator learning for partial differential equations \cite{sojitra2026fedonet,abid2025spectral,abid2026wlno}. These developments are relevant to turbulent reconstruction, although neither an operator architecture nor a physics-based loss alone resolves the fundamental information loss caused by severe downsampling. The choice of observations, training distribution, and validation diagnostics remains central.

Despite these advances, a substantial portion of the
existing literature on data-driven turbulent
super-resolution relies on supervised learning
frameworks trained using paired low- and
high-resolution flow fields
\cite{fukami2019superresolution,
fukami2021spatiotemporal,
yousif2022gan,guastoni2021convolutional}.
In such approaches, high-resolution reference
fields are typically obtained from direct
numerical simulations or experimental
measurements, while corresponding
low-resolution observations are generated
through spatial filtering, downsampling,
or measurement-based degradation.
Although these frameworks have demonstrated
considerable success in reconstructing
unresolved turbulent structures, their
performance depends on the availability
and representativeness of the training data.
Constructing sufficiently diverse
high-fidelity datasets can require
repeated computationally expensive
simulations, extensive experimental
measurements, or sampling across multiple
flow states and parameter configurations
\cite{duraisamy2019turbulence,
brunton2020machine,fukami2023survey}.
This requirement presents a practical
challenge, particularly when the
objective of super-resolution is itself
to mitigate the scarcity of
high-resolution flow information.
The difficulty becomes especially
pronounced for previously unexplored
flow regimes, costly Reynolds-number
or parameter sweeps, and applications
in which only one or a few
high-resolution realizations are
available.

Homogeneous turbulence suggests an alternative source of training information. Statistical homogeneity implies translation-invariant statistics, so patches extracted at different locations probe the same underlying statistical process \cite{batchelor1953theory,pope2000turbulent}. A single sufficiently large realization contains many localized vortical, strain-dominated, and gradient-rich configurations across scales. These patches are not independent samples, and homogeneity does not imply that one snapshot contains every relevant state. Nevertheless, internal spatial sampling can expose a learning algorithm to numerous coarse-to-fine configurations without requiring additional global snapshots. Related ideas have a precedent in single-image internal statistics and zero-shot super-resolution \cite{zontak2011internal,shocher2018zeroshot}, where examples extracted from an individual image are used to train an image-specific reconstruction model.

Fukami and Taira \cite{fukami2024single} demonstrated the feasibility of single-snapshot learning for turbulent super-resolution by training a convolutional neural network on localized subdomains extracted from a single high-resolution flow realization. Their study demonstrated generalization to previously unseen turbulent states, including two-dimensional decaying isotropic turbulence and turbulent channel flow, highlighting the potential of exploiting internal spatial information under severe data scarcity. However, their analysis primarily focused on reconstruction within individual computational subdomains, leaving the implications of assembling independently predicted regions into complete high-resolution turbulent fields less extensively explored. In particular, maintaining spatial continuity across reconstructed subdomains while preserving the global multiscale characteristics of turbulence presents an additional challenge. Motivated by this distinction, the present study investigates full-domain extreme super-resolution through overlapping patchwise prediction and Hann-window-weighted overlap-add reconstruction, considering enlargement factors of $\times32$ and $\times64$.

At these extreme enlargement factors, the low-resolution observations retain only a limited range of resolved wavenumbers, rendering the recovery of fine-scale turbulent structures fundamentally underdetermined. The objective is therefore not merely to reproduce plausible high-frequency fluctuations but to determine whether coarse-to-fine relationships learned from a single realization can recover structures correlated with previously unseen turbulent fields. To assess this capability, we examine physical-space reconstruction errors, coherent vortical structures, turbulent energy spectra, vorticity probability distributions, and scale-dependent Fourier coherence. The latter provides a complementary measure of realization-specific cross-spectral correspondence, since agreement in angle-averaged energy spectra alone does not establish the recovery of the Fourier organization of an individual turbulent realization.

Recent advances in deep-learning-based super-resolution
have introduced increasingly sophisticated architectures
for recovering fine-scale spatial information from
low-resolution observations. Residual convolutional
frameworks, including 
residual dense networks (RDN) \cite{zhang2018residual},
and residual channel attention networks (RCAN)
\cite{zhang2018image}, have demonstrated the effectiveness
of deep feature extraction and residual learning.
Subsequent developments have incorporated adversarial
learning, as exemplified by ESRGAN
\cite{wang2018esrgan}, and transformer-based architectures,
including SwinIR \cite{liang2021swinir} and hybrid
attention transformers (HAT) \cite{chen2025hat},
to further improve the reconstruction of fine-scale
image features. In this work, we investigate extreme super-resolution of two-dimensional decaying homogeneous isotropic turbulence using a wide-activation deep super-resolution (WDSR) network \cite{yu2018wide}. Training examples are generated by extracting overlapping patches from a single high-resolution vorticity realization and constructing corresponding coarse-to-fine pairs. For inference on unseen realizations, overlapping predictions are assembled using Hann-window-weighted normalized overlap-add to reduce patch-boundary artifacts. We consider $\times32$ and $\times64$ reconstruction and assess relative errors, vortical structure, energy spectra, vorticity probability distributions, and scale-dependent Fourier coherence. We also examine overlap-add variants and the sensitivity of reconstruction to perturbations in the coarse observations.

To contextualize the internal-learning formulation, we compare it with conventional full-field WDSR models trained using $1$, $25$, $50$, and $100$ high-resolution snapshots. This comparison tests the practical effectiveness of two learning strategies: increasing the number of complete realizations versus extracting local examples from one realization. Because the strategies differ in training-data presentation and inference assembly as well as snapshot count, their comparison is not interpreted as a controlled estimate of the isolated effect of dataset size. Instead, it measures the performance of the respective end-to-end formulations under the specified protocols.

The central contribution of this study is thus an assessment of the capabilities and limits of single-realization internal learning at extreme enlargement factors, rather than the introduction of single-snapshot learning itself. By combining $\times32$ and $\times64$ reconstruction with physical-space diagnostics, spectral statistics, Fourier coherence, and a comparison with multi-snapshot full-field training, we examine whether the information distributed across one homogeneous turbulent realization can provide a useful transferable prior when high-resolution training data are severely limited. The results aim to clarify both the promise and the inherent non-uniqueness of data-efficient turbulent super-resolution.

The remainder of this paper is organized as follows.
Section~\ref{sec:approach} presents the proposed single-realization
super-resolution framework, including the WDSR architecture,
overlapping patch extraction, Hann-window-weighted reconstruction,
and evaluation methodology.
Section~\ref{sec:results} discusses the numerical results for
two-dimensional homogeneous isotropic turbulence at extreme
super-resolution factors of $\times 32$ and $\times 64$,
including reconstruction accuracy, spectral and statistical
fidelity, robustness analyses, and comparisons with conventional
multi-realization learning.
Finally, Section~\ref{sec:conclusions} summarizes the principal
findings, limitations, and implications for data-efficient
turbulent flow reconstruction.

\section{Approach}
\label{sec:approach}

\subsection{Single-realization learning framework}
\label{sec:single_realization}

The proposed framework exploits the spatial redundancy of homogeneous
turbulence to construct training examples from a single high-resolution
realization. Under statistical homogeneity, spatially separated regions
share the same underlying statistical properties, allowing localized
coarse-to-fine relationships to be sampled without requiring additional
flow realizations. The objective is to learn these relationships from
one turbulent field and assess their transferability to previously
unseen realizations.

Let $q^{\mathrm{HR}}(\boldsymbol{x})$, with
$\boldsymbol{x}\in\Omega$, denote a single high-resolution
turbulent field defined over the spatial domain $\Omega$.
In the present study, $q$ represents the vorticity field of
two-dimensional homogeneous isotropic turbulence.
A collection of overlapping high-resolution patches is
extracted according to
\begin{equation}
q_i^{\mathrm{HR}}
=
\mathcal{P}_i
\left[q^{\mathrm{HR}}\right]
\in
\mathbb{R}^{N_p\times N_p},
\qquad
i=1,\ldots,N_{\mathrm{tr}},
\label{eq:patch_extraction}
\end{equation}
where $\mathcal{P}_i$ denotes the spatial extraction operator,
$N_p$ is the patch width, and $N_{\mathrm{tr}}$ is the total
number of training patches.

The corresponding low-resolution patches are obtained
through a prescribed spatial degradation operator,
\begin{equation}
q_i^{\mathrm{LR}}
=
\mathcal{D}_r
\left[q_i^{\mathrm{HR}}\right],
\label{eq:downsampling}
\end{equation}
where $\mathcal{D}_r$ denotes spatial coarsening by a
linear factor $r$. We consider the extreme super-resolution
factors $r=32$ and $r=64$, corresponding to reductions
of $1024$ and $4096$, respectively, in the number of
spatial samples.

The resulting training pairs define the supervised
learning problem
\begin{equation}
\mathcal{G}_{\theta}:
q_i^{\mathrm{LR}}
\longmapsto
q_i^{\mathrm{HR}},
\label{eq:sr_mapping}
\end{equation}
where $\mathcal{G}_{\theta}$ is the parameterized
super-resolution mapping and $\theta$ denotes its
trainable parameters. All training pairs
$\{q_i^{\mathrm{LR}},q_i^{\mathrm{HR}}\}$ are derived
exclusively from the same high-resolution realization,
without introducing additional snapshots or external
turbulent datasets.

Rather than learning the global spatial arrangement of
a particular realization, the network is trained to
approximate recurring relationships between resolved
coarse-scale features and their associated fine-scale
structures. Its ability to generalize is subsequently
evaluated on turbulent realizations excluded from training,
providing a direct assessment of the transferability of
the learned multiscale relationships.
\subsection{WDSR super-resolution model}
\label{sec:wdsr}

\begin{figure*}[ht!]
    \centering
    \includegraphics[width=\textwidth]{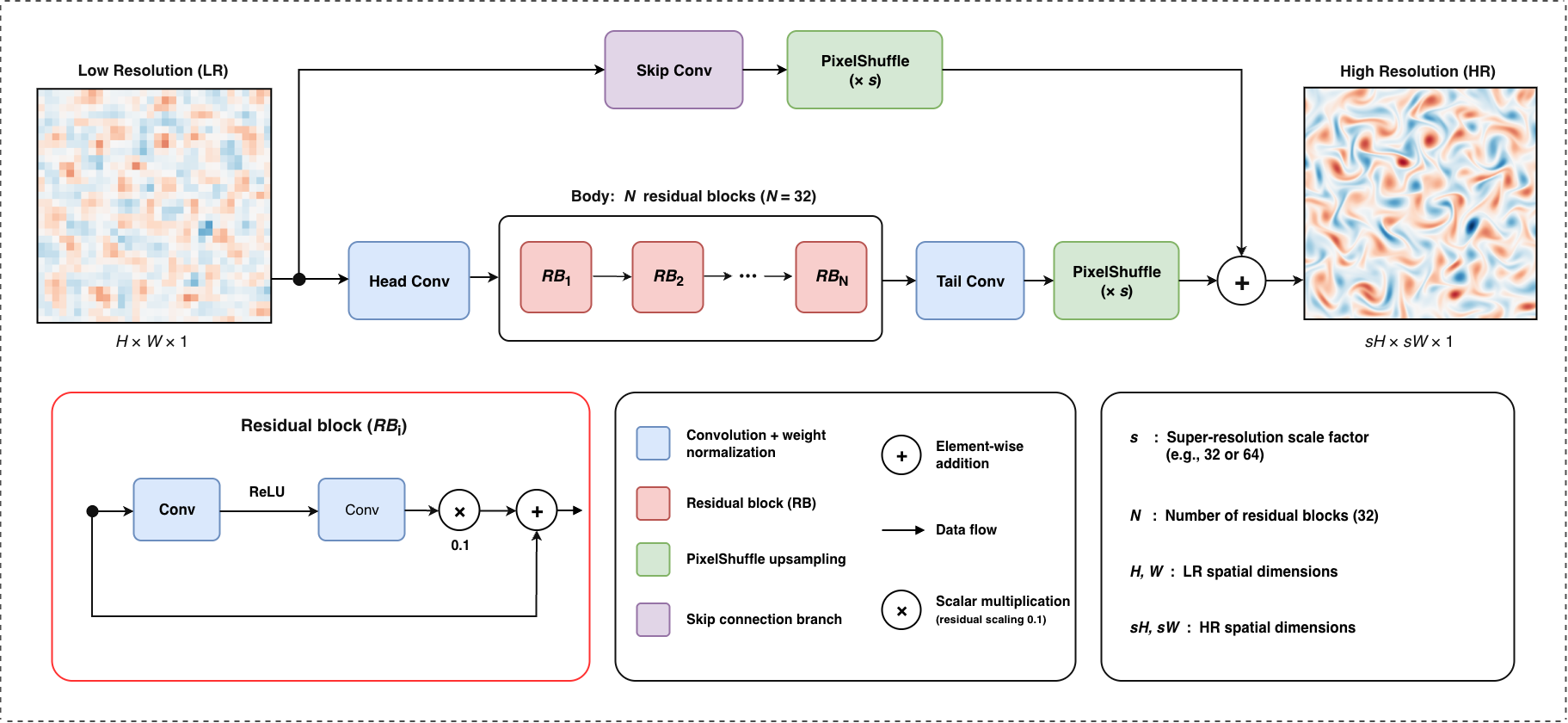}
    \caption{
    Architecture of the WDSR network employed for extreme-scale turbulent
    super-resolution.
    }
    \label{fig:wdsr_framework}
\end{figure*}

The nonlinear mapping introduced in Eq.~(\ref{eq:sr_mapping}) is approximated
using a wide-activation deep super-resolution (WDSR) network. The architecture
employed in the present study is illustrated in
Fig.~\ref{fig:wdsr_framework}. Importantly, the network operates directly on
the low-resolution input and performs the required increase in spatial
resolution internally through learned upsampling. No interpolation to the
target resolution is performed before the field is supplied to the network.

For an input patch
$q_i^{\mathrm{LR}}\in\mathbb{R}^{H\times W}$, the network produces
\begin{equation}
\widehat{q}_i^{\mathrm{HR}}
=
\mathcal{G}_{\theta}
\left(
q_i^{\mathrm{LR}}
\right),
\qquad
\widehat{q}_i^{\mathrm{HR}}
\in
\mathbb{R}^{rH\times rW},
\label{eq:wdsr_mapping}
\end{equation}
where $r$ denotes the prescribed linear super-resolution factor.

As shown in Fig.~\ref{fig:wdsr_framework}, the architecture consists of a
main nonlinear reconstruction branch and a parallel skip branch. The main
branch first projects the low-resolution input into a latent feature space
through an initial convolutional layer. The resulting feature representation
is subsequently processed by a sequence of $N=32$ residual blocks. Within
each block, two convolutional transformations are employed, with a rectified
linear unit (ReLU) activation applied between them. The residual contribution
is scaled by a factor of $0.1$ before being added to the block input. This
residual scaling limits the magnitude of successive residual updates and
facilitates stable optimization of the deep reconstruction network.

Following the residual body, a tail convolution generates the feature
channels required for learned spatial upsampling. A PixelShuffle operation
then rearranges these channels into the desired high-resolution spatial
representation. In parallel, the skip branch applies a convolutional
transformation and PixelShuffle operation directly to the low-resolution
input. The outputs of the nonlinear reconstruction and skip branches are
combined through elementwise addition to obtain the final high-resolution
prediction.

The residual formulation provides a useful decomposition for turbulent
super-resolution. The coarse observation contains the dominant resolved
spatial organization, whereas the principal reconstruction task concerns the
information removed by spatial coarsening, including sharp gradients,
localized vortical structures, and high-wavenumber fluctuations. At the
extreme scale factors considered here, however, the unresolved contribution
constitutes a substantial fraction of the spatial information contained in
the reference field. The reconstruction problem therefore extends well beyond
conventional interpolation and requires the network to infer fine-scale
structure using statistical relationships acquired during training.

The network parameters are obtained by minimizing the mean-squared
reconstruction loss over the $N_{\mathrm{tr}}$ training patches,
\begin{equation}
\mathcal{L}(\theta)
=
\frac{1}{N_{\mathrm{tr}}}
\sum_{i=1}^{N_{\mathrm{tr}}}
\left\|
\mathcal{G}_{\theta}
\left(
q_i^{\mathrm{LR}}
\right)
-
q_i^{\mathrm{HR}}
\right\|_2^2.
\label{eq:training_loss}
\end{equation}

It is emphasized that WDSR itself is not introduced here as a new
super-resolution architecture. Rather, it provides a consistent and effective
reconstruction model through which the principal hypothesis of the present
work can be examined: whether the internal spatial statistics contained within
a single turbulent realization provide sufficient information to support
extreme-scale reconstruction of previously unseen turbulent states.

\subsection{Patchwise inference and overlap-add reconstruction}
\label{sec:patch_reconstruction}

\begin{figure*}[ht!]
    \centering
    \includegraphics[width=\textwidth]
    {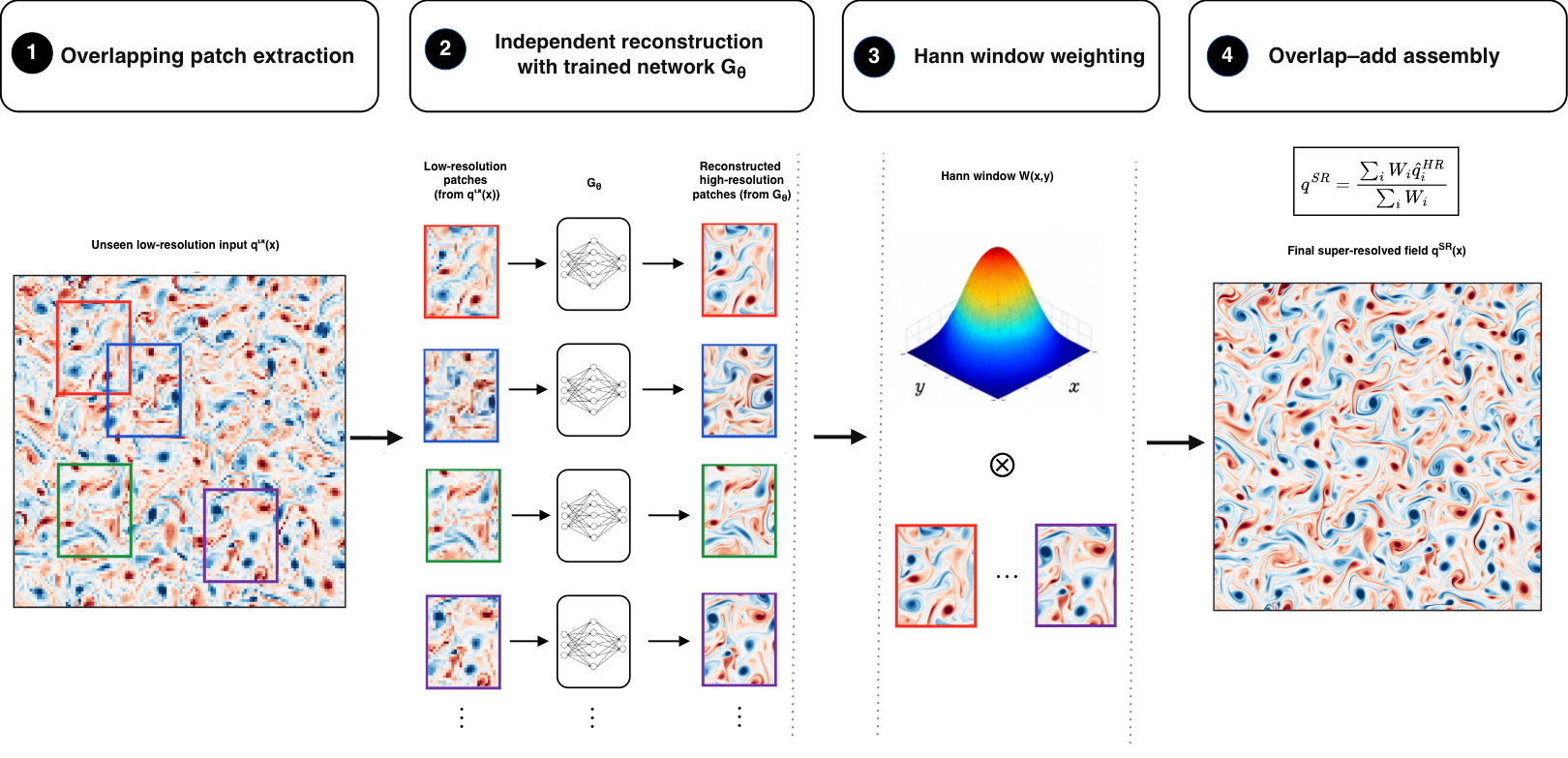}
    \caption{
    Patchwise reconstruction and normalized overlap-add assembly of an unseen
    turbulent realization. 
    }
    \label{fig:patch_reconstruction}
\end{figure*}

Following training, the learned mapping is applied to turbulent realizations
that are entirely unseen during optimization. Reconstruction is performed
locally rather than through a single full-field network evaluation. The
procedure, illustrated schematically in
Fig.~\ref{fig:patch_reconstruction}, consists of four stages:
(i) overlapping patch extraction from the unseen low-resolution field,
(ii) independent super-resolution of each patch using the trained WDSR model,
(iii) spatial weighting of the reconstructed patches using a two-dimensional
Hann window, and
(iv) normalized overlap-add assembly of the complete super-resolved field.

Let $q_i^{\mathrm{LR}}$ denote the $i$th overlapping patch extracted from an
unseen low-resolution realization. Each patch is independently mapped to its
high-resolution representation according to
\begin{equation}
\widehat{q}_i^{\mathrm{HR}}
=
\mathcal{G}_{\theta}
\left(
q_i^{\mathrm{LR}}
\right).
\label{eq:patch_inference}
\end{equation}

Overlapping extraction results in multiple predictions for spatial locations
shared by neighboring patches. Direct concatenation of independently
reconstructed patches can produce localized discontinuities or stitching
artefacts because predictions near patch boundaries may be less reliable than
those in the patch interior. To suppress these effects, each reconstructed
patch is modulated using a separable two-dimensional Hann window,
\begin{equation}
W(m,n)
=
w(m)w(n),
\label{eq:hann2d}
\end{equation}
where
\begin{equation}
w(j)
=
\frac{1}{2}
\left[
1-
\cos
\left(
\frac{2\pi j}{N_p-1}
\right)
\right],
\qquad
j=0,\ldots,N_p-1.
\label{eq:hann1d}
\end{equation}

Here, $N_p$ denotes the spatial dimension of the reconstructed
high-resolution patch. The Hann window assigns greater weight to the interior
of each prediction and smoothly reduces its contribution toward the patch
boundaries.

The weighted contribution of the $i$th reconstructed patch is defined as
\begin{equation}
\widetilde{q}_i^{\mathrm{HR}}(\boldsymbol{x})
=
W_i(\boldsymbol{x})
\widehat{q}_i^{\mathrm{HR}}(\boldsymbol{x}),
\label{eq:weighted_patch}
\end{equation}
where $W_i$ denotes the Hann window positioned at the spatial location
corresponding to the $i$th patch. The complete super-resolved field is then
obtained through normalized overlap-add reconstruction,
\begin{equation}
q^{\mathrm{SR}}(\boldsymbol{x})
=
\frac{
\displaystyle
\sum_i
W_i(\boldsymbol{x})
\widehat{q}_i^{\mathrm{HR}}(\boldsymbol{x})
}{
\displaystyle
\sum_i
W_i(\boldsymbol{x})
}.
\label{eq:overlap_add}
\end{equation}

The normalization in Eq.~(\ref{eq:overlap_add}) accounts for both the number
and relative weights of overlapping predictions contributing to each spatial
location. Neighboring reconstructions are therefore blended continuously
rather than joined through hard patch boundaries. This treatment becomes
particularly important at extreme super-resolution factors, where localized
patch-boundary inconsistencies can become pronounced after large increases in
spatial resolution.

It should be emphasized that the overlap-add procedure introduces neither
additional trainable parameters nor information from the high-resolution
reference during inference. It acts solely as a deterministic spatial blending
operator for assembling independently generated local predictions into a
globally continuous super-resolved field.

\subsection{Comparison with multi-realization learning}
\label{sec:data_scaling}

To assess the effectiveness of exploiting spatial information
within a single turbulent realization, we compare the proposed
patch-based framework with conventional WDSR models trained
on progressively larger collections of high-resolution
snapshots. Specifically, separate models are trained using
$N_s\in\{1,25,50,100\}$ complete turbulent realizations
without overlapping patch extraction or patchwise
overlap-add reconstruction.

The comparison examines two distinct learning strategies:
extracting localized coarse-to-fine relationships from a
single high-resolution realization and learning from an
increasing number of complete turbulent snapshots.
All models are evaluated on the same previously unseen
turbulent fields using identical reconstruction metrics
and physical diagnostics.

Because these strategies differ in both training-data
presentation and inference procedure, the comparison is
not intended to isolate the effect of training-set size.
Instead, it evaluates the relative performance of the
complete learning and reconstruction formulations under
the specified protocols, providing a practical assessment
of their effectiveness when high-resolution turbulent
training data are limited.

\subsection{Evaluation methodology}
\label{sec:evaluation}

Reconstruction accuracy is assessed using complementary
physical-space, spectral, and statistical diagnostics to
evaluate both pointwise fidelity and the recovery of
multiscale turbulent structures. For a reconstructed field
$q^{\mathrm{SR}}$ and its high-resolution reference
$q^{\mathrm{HR}}$, the relative $L_2$ error is defined as
\begin{equation}
\epsilon_{L_2}
=
\frac{
\left\|
q^{\mathrm{SR}}-q^{\mathrm{HR}}
\right\|_2
}{
\left\|
q^{\mathrm{HR}}
\right\|_2
}.
\label{eq:relative_l2}
\end{equation}
Root-mean-square error (RMSE), mean absolute error (MAE),
and peak signal-to-noise ratio (PSNR) are also used to
quantify reconstruction accuracy. Spatial error
distributions and comparisons of the $Q$-criterion
provide additional information about localized
reconstruction errors and the recovery of flow-gradient
structures.

Spectral fidelity is evaluated using angle-averaged
energy spectra, with particular attention to the
wavenumber range beyond the low-resolution Nyquist
cutoff. Scale-dependent Fourier coherence is additionally
examined to assess normalized cross-spectral
correspondence between the reconstructed and reference
fields. The distinction between spectral-energy agreement
and realization-specific Fourier correspondence is
important because similar energy distributions do not
necessarily imply recovery of the same fine-scale
spatial organization.

The statistical properties of the reconstructed
turbulence are examined through vorticity probability
density functions, including their tails associated
with intense fluctuations. Together, these diagnostics
provide complementary measures of reconstruction
quality across physical and spectral representations.
All learning strategies are evaluated using the same
unseen turbulent realizations and consistent diagnostic
definitions to ensure comparability of the reported
results.
\section{Results}
\label{sec:results}

\subsection{Two-dimensional homogeneous isotropic turbulence}
\label{sec:hit_results}

We evaluate the proposed single-realization learning framework using
two-dimensional decaying homogeneous isotropic turbulence at
$Re=1000$. The flow evolution is governed by the incompressible
vorticity transport equation

\begin{equation}
\frac{\partial \omega}{\partial t}
+
J(\omega,\psi)
=
\frac{1}{Re}\nabla^2\omega,
\label{eq:vorticity_results}
\end{equation}

where $\omega$ denotes the vorticity, $\psi$ is the streamfunction, and

\begin{equation}
J(\omega,\psi)
=
\frac{\partial \omega}{\partial x}
\frac{\partial \psi}{\partial y}
-
\frac{\partial \omega}{\partial y}
\frac{\partial \psi}{\partial x}
\end{equation}

is the nonlinear Jacobian operator. The streamfunction is obtained from
the Poisson equation

\begin{equation}
\nabla^2\psi=-\omega,
\label{eq:poisson_results}
\end{equation}

with the corresponding velocity components given by

\begin{equation}
u=\frac{\partial\psi}{\partial y},
\qquad
v=-\frac{\partial\psi}{\partial x}.
\label{eq:velocity_results}
\end{equation}

The DNS data are represented on a $2048\times2048$ spatial grid in a
doubly periodic domain of size $[0,2\pi]\times[0,2\pi]$ following the numerical formulation described
in Ref.~\cite{pawar2023frame} . The resulting
flow fields contain a broad distribution of coherent vortical
structures, elongated vorticity filaments, and localized gradients
spanning a wide range of spatial scales. The repeated occurrence of
these structures throughout the homogeneous domain provides the
internal spatial information exploited by the proposed learning
framework.

\begin{figure}[ht!]
    \centering
    \includegraphics[width=0.65\linewidth]{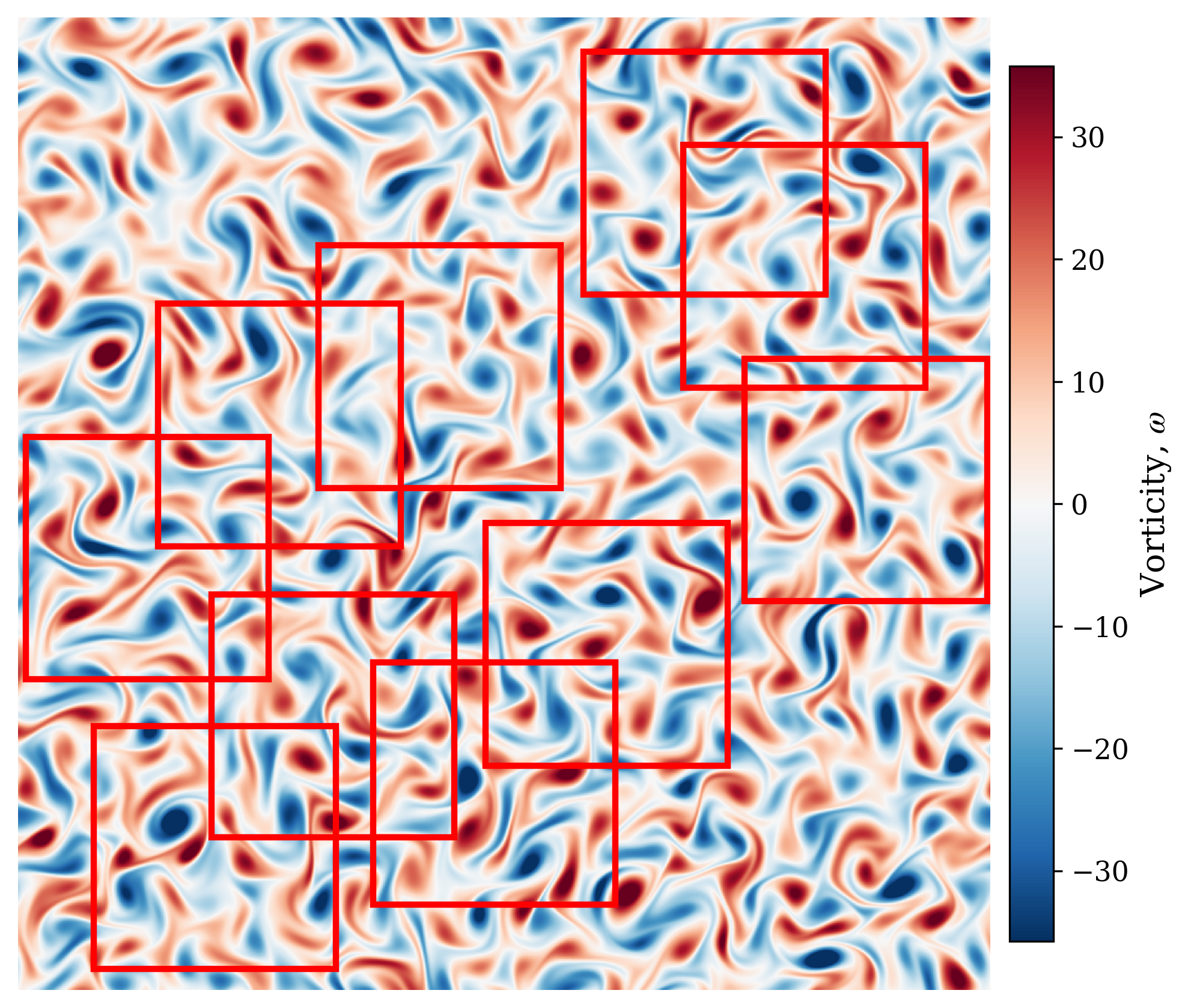}
    \caption{
    High-resolution vorticity realization of two-dimensional decaying
    homogeneous isotropic turbulence at $Re=1000$ used to construct the
    single-realization training dataset. Red boxes illustrate representative
    overlapping spatial patches extracted from this single realization for
    training the WDSR model.
    }
    \label{fig:training_snapshot}
\end{figure}

For the single-realization model, the complete training dataset is
constructed exclusively from one high-resolution vorticity realization.
Figure~\ref{fig:training_snapshot} shows this realization together with
representative spatial regions used during patch extraction. Rather than
treating the $2048\times2048$ field as a single training sample,
overlapping spatial regions are extracted to expose the network to
localized manifestations of the multiscale turbulent organization
distributed throughout the domain. Importantly, every training sample
originates from the same instantaneous flow field; no additional
turbulent snapshots or temporal states are introduced during training.

Generalization is evaluated on turbulent realizations that are entirely
excluded from training. The present investigation considers two extreme super-resolution factors, $r=32$ and $r=64$. For the
$2048\times2048$ reference resolution, the corresponding reconstruction
problems are

\begin{equation}
64\times64
\overset{\times32}{\longrightarrow}
2048\times2048,
\label{eq:x32_mapping}
\end{equation}

and

\begin{equation}
32\times32
\overset{\times64}{\longrightarrow}
2048\times2048.
\label{eq:x64_mapping}
\end{equation}

Unless otherwise stated, all WDSR results presented below correspond to
the model trained using the single high-resolution realization shown in
Fig.~\ref{fig:training_snapshot} and subsequently evaluated on unseen
turbulent fields. Bicubic interpolation is employed as the conventional
nonlearned reconstruction baseline.

\subsection{Extreme super-resolution from a single realization}
\label{sec:extreme_sr}

\subsubsection{Super-resolution by a factor of $\times32$}
\label{sec:x32_results}

\begin{figure*}[ht!]
    \centering
    \includegraphics[width=\textwidth]{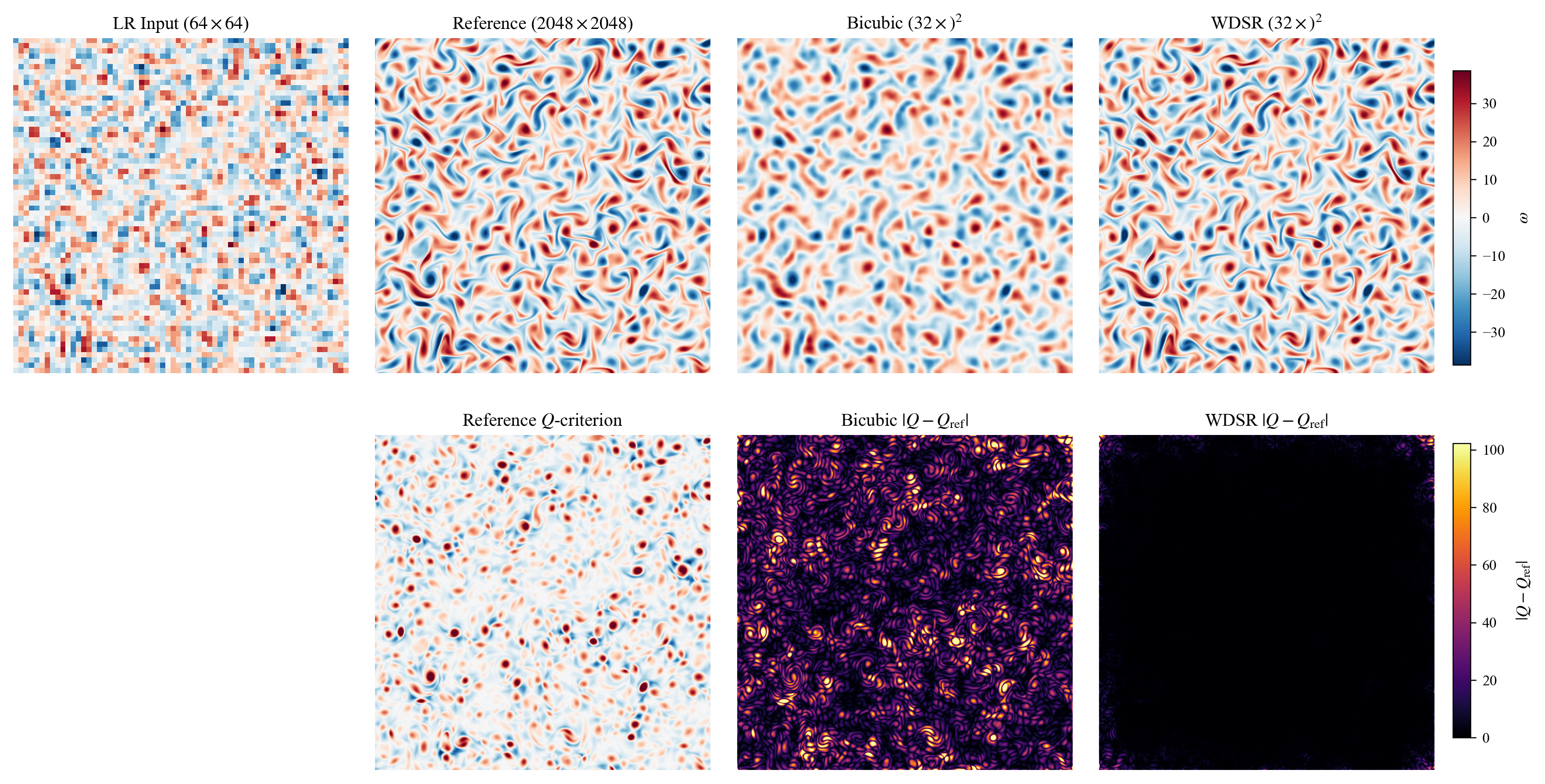}
    \caption{
    Extreme $\times32$ super-resolution of an unseen vorticity
    realization at $Re=1000$. The upper row shows, from left to right,
    the $64\times64$ low-resolution input, $2048\times2048$ DNS
    reference, bicubic interpolation, and WDSR reconstruction obtained
    using the single-realization overlapping-patch framework. The lower
    row shows the reference $Q$-criterion and the corresponding absolute
    $Q$-criterion errors for bicubic interpolation and WDSR.
    }
    \label{fig:sr32_fields}
\end{figure*}

We first consider the $\times32$ reconstruction problem, in which a
$64\times64$ observation is used to reconstruct the corresponding
$2048\times2048$ vorticity field. This represents a reduction by three
orders of magnitude in the number of spatial samples available to the
reconstruction model. Figure~\ref{fig:sr32_fields} compares the
low-resolution observation, DNS reference, bicubic reconstruction, and
single-realization WDSR prediction for an unseen turbulent realization.

The severity of the spatial information loss is apparent from the
low-resolution field in Fig.~\ref{fig:sr32_fields}. At a resolution of
only $64\times64$, many of the thin vorticity filaments, localized
extrema, and sharp gradients present in the DNS field are no longer
explicitly resolved. Bicubic interpolation restores the target grid
dimensions but produces a visibly smoothed representation. Although the
largest spatial features are retained, much of the filamentary
fine-scale organization is attenuated or absent.

In contrast, the WDSR reconstruction exhibits close
qualitative agreement with the DNS reference throughout
the computational domain. The model recovers thin vortical
filaments, localized regions of intense vorticity, and
substantially more of the multiscale spatial organization
than bicubic interpolation. This improvement is particularly
notable because the reconstructed realization is entirely
absent from the training dataset. The qualitative
observations are supported by the quantitative reconstruction
metrics. Bicubic interpolation
produces a relative $L_2$ error of $36.64\%$, whereas
the proposed single-realization WDSR framework reduces
this error to $4.25\%$, corresponding to an approximately
$88.4\%$ reduction in relative error. Similarly, the
root-mean-square error decreases from $5.0773$ to
$0.5893$, while the mean absolute error decreases from
$3.8709$ to $0.1772$. The peak signal-to-noise ratio
increases from $26.78$~dB for bicubic interpolation
to $45.48$~dB for WDSR.

The lower row of Fig.~\ref{fig:sr32_fields} provides
a complementary assessment based on the $Q$-criterion,
which characterizes the local balance between
rotation- and strain-dominated regions of the flow.
Bicubic interpolation produces widespread errors
concentrated around coherent vortical structures
and regions containing strong spatial gradients.
The WDSR reconstruction substantially suppresses
both the magnitude and spatial extent of these
errors, indicating improved recovery of the
local flow-gradient organization. Consequently,
the improvements are not restricted to the
pointwise accuracy of the reconstructed vorticity
field but also extend to derived quantities
associated with coherent turbulent structures.

\begin{figure}[ht!]
    \centering
    \includegraphics[width=0.75\linewidth]{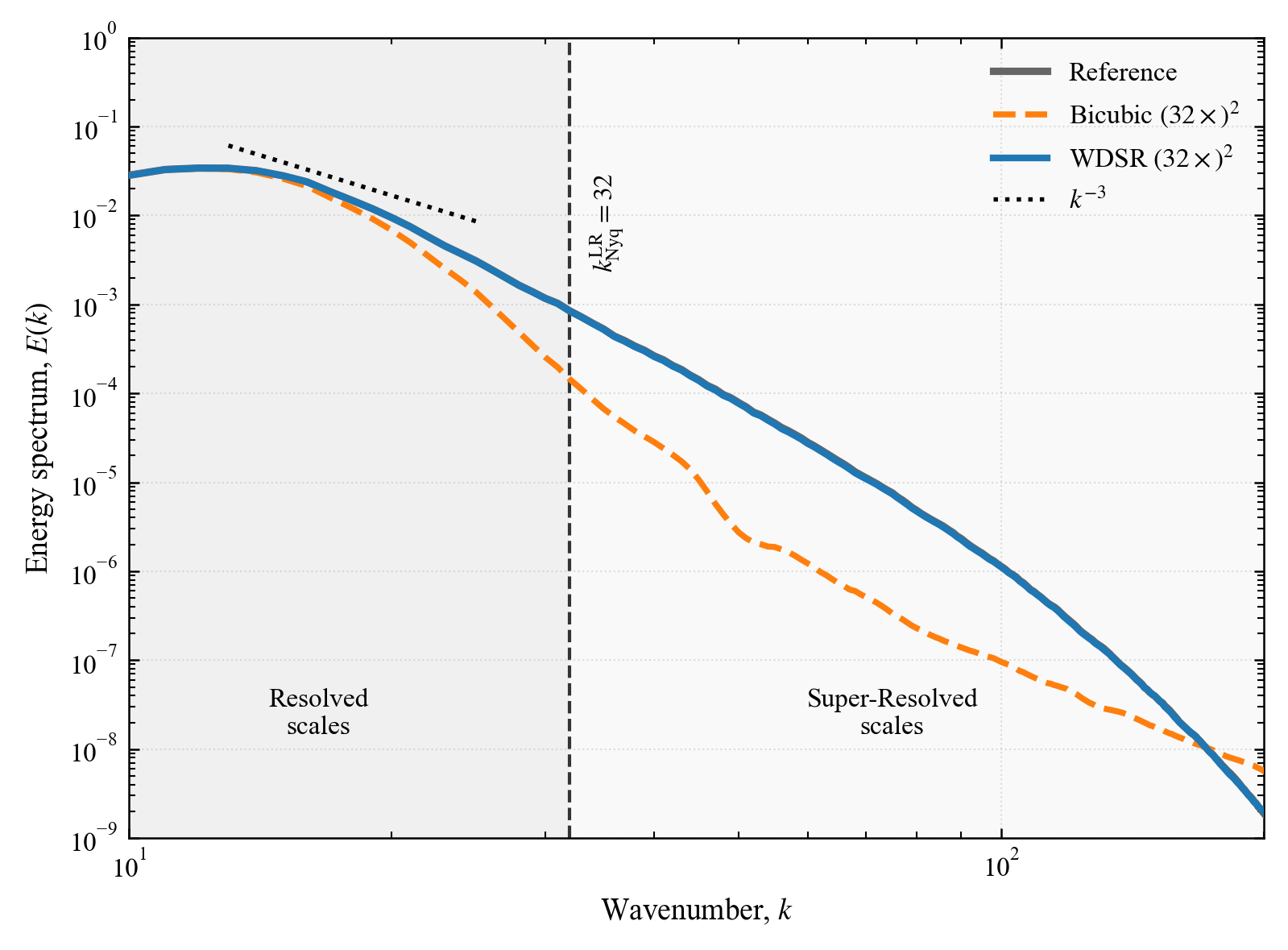}
    \caption{
    Angle-averaged energy spectrum for the $\times32$ reconstruction of
    an unseen turbulent realization at $Re=1000$. The DNS reference is
    compared with bicubic interpolation and the single-realization WDSR
    reconstruction. The vertical dashed line denotes the low-resolution
    Nyquist wavenumber, $k_{\mathrm{Nyq}}^{\mathrm{LR}}=32$, separating
    scales represented by the $64\times64$ input from scales inferred
    through super-resolution. The $k^{-3}$ line is included as a
    reference scaling.
    }
    \label{fig:spectrum32}
\end{figure}

A more stringent assessment is obtained in spectral space. Figure
\ref{fig:spectrum32} compares the angle-averaged energy spectra obtained
from the DNS reference, bicubic interpolation, and WDSR reconstruction.
The vertical dashed line indicates the low-resolution Nyquist
wavenumber, $k_{\mathrm{Nyq}}^{\mathrm{LR}}=32$, thereby separating
scales that are directly represented in the $64\times64$ observation
from those that must be inferred through super-resolution.

The WDSR spectrum closely follows the DNS reference throughout the
low-wavenumber range and, importantly, remains in close agreement well
beyond $k_{\mathrm{Nyq}}^{\mathrm{LR}}$. In contrast, bicubic
interpolation progressively attenuates the spectral energy and departs
strongly from the reference as the wavenumber increases. This behavior
is expected because interpolation can increase the number of grid
points but cannot reconstruct the turbulent spectral content removed
during spatial coarsening.

The distinction becomes particularly clear for
$k>k_{\mathrm{Nyq}}^{\mathrm{LR}}$, where the corresponding spatial
scales are absent from the original low-resolution observation. The
WDSR model reconstructs a spectral distribution that remains close to
the DNS reference over a broad portion of this super-resolved range.
Only toward the highest wavenumbers does the reconstructed spectrum
begin to decay more rapidly than the reference. The result therefore
indicates that the learned model recovers physically meaningful
multiscale content substantially beyond the spectral bandwidth supplied
to it, while also revealing a finite limit to the smallest scales that
can be reconstructed accurately.

\begin{figure}[ht!]
    \centering
    \includegraphics[width=0.75\linewidth]{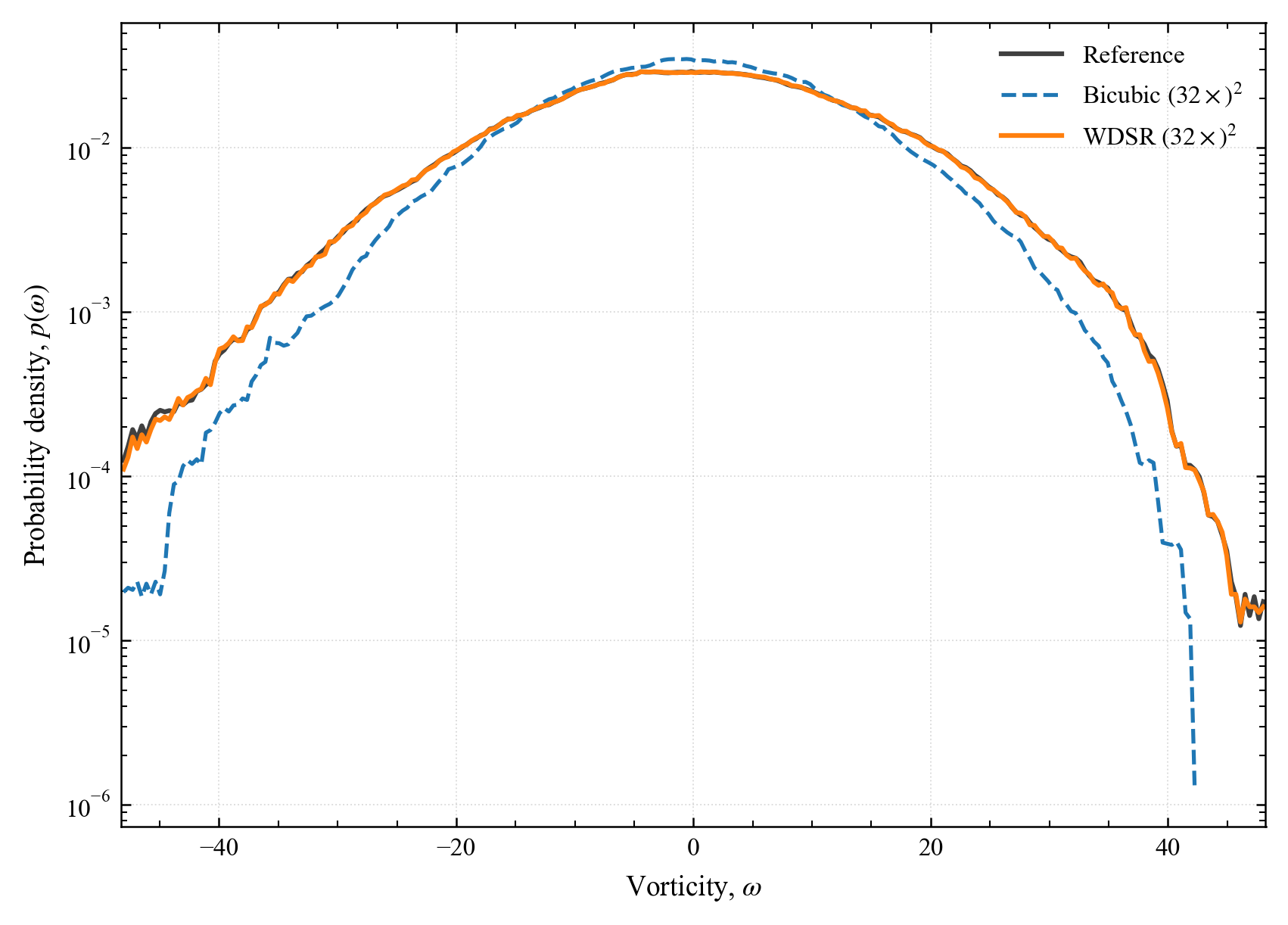}
    \caption{
    Probability density function of vorticity for the $\times32$
    reconstruction at $Re=1000$. The DNS reference is compared with
    bicubic interpolation and the single-realization WDSR
    reconstruction.
    }
    \label{fig:pdf32}
\end{figure}

The recovery of one-point vorticity statistics is examined in
Fig.~\ref{fig:pdf32}, which compares the probability density functions
(PDFs) of the DNS, bicubic, and WDSR fields.

The WDSR distribution is nearly coincident with the reference PDF over
most of the vorticity range. In particular, the model reproduces both
the width of the central distribution and the low-probability tails
associated with large positive and negative vorticity fluctuations.
These tails are important because they correspond to relatively rare,
high-intensity events that are strongly affected by spatial smoothing.

Bicubic interpolation, by comparison, produces a substantially narrower
distribution. The probability of large-amplitude vorticity is strongly
suppressed, consistent with both the loss of intense localized
structures observed in physical space and the attenuation of
high-wavenumber content observed in Fig.~\ref{fig:spectrum32}. The
agreement of the WDSR prediction with the DNS across physical,
spectral, and statistical diagnostics therefore provides substantially
stronger evidence of reconstruction fidelity than visual similarity
alone.

\subsubsection{Super-resolution by a factor of $\times64$}
\label{sec:x64_results}

\begin{figure*}[ht!]
    \centering
    \includegraphics[width=\textwidth]{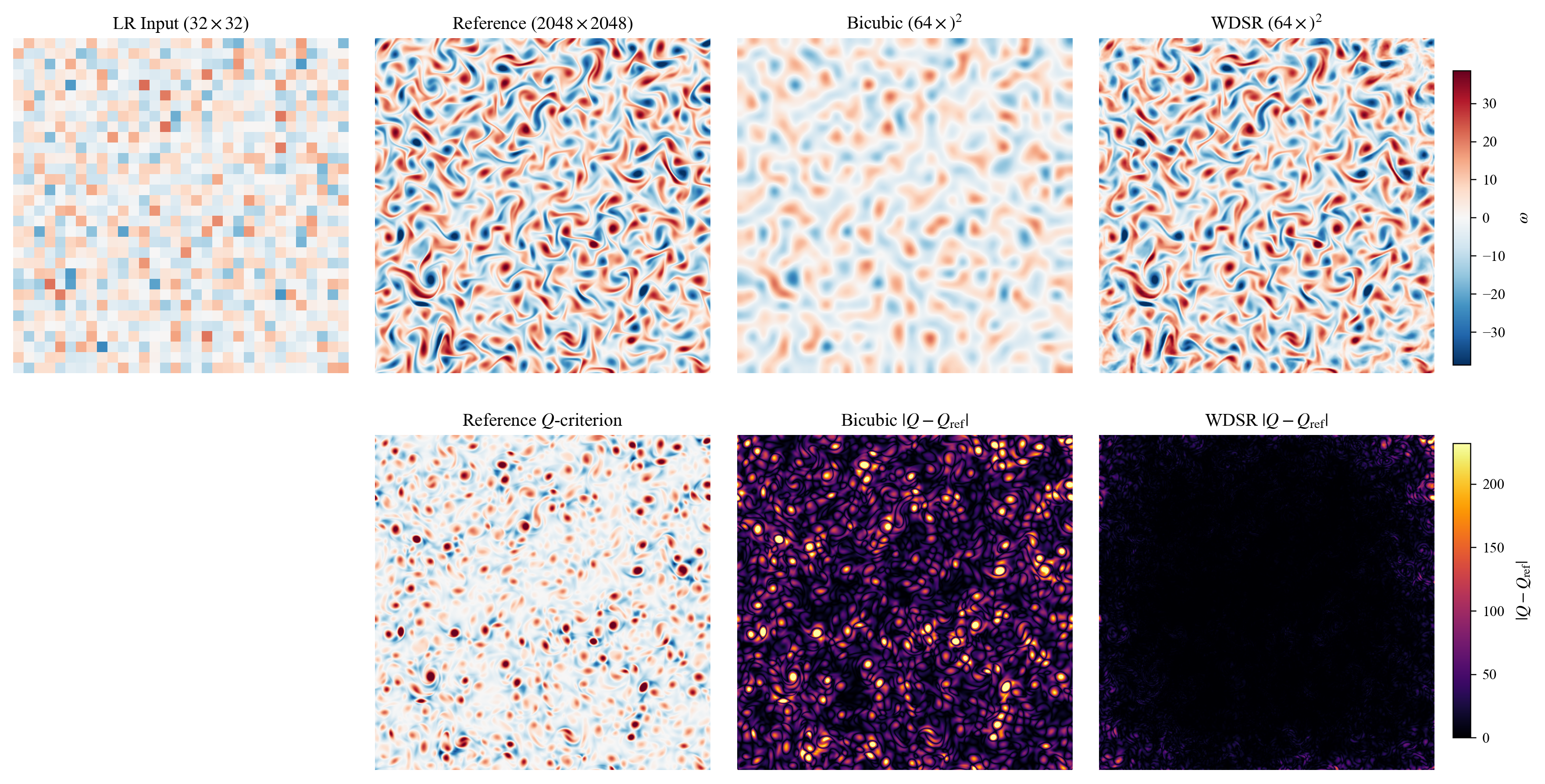}
    \caption{
    Extreme $\times64$ super-resolution of an unseen
    vorticity realization at $Re=1000$. The upper row
    shows, from left to right, the $32\times32$
    low-resolution input, $2048\times2048$ DNS
    reference, bicubic interpolation, and WDSR
    reconstruction obtained using the single-realization
    overlapping-patch framework. The lower row shows
    the reference $Q$-criterion and the corresponding
    absolute $Q$-criterion errors for bicubic
    interpolation and WDSR.
    }
    \label{fig:sr64_fields}
\end{figure*}

We next consider the more challenging $\times64$
super-resolution problem, in which a $2048\times2048$
DNS vorticity field is reconstructed from a
$32\times32$ observation containing only $1/4096$
of the spatial samples in the reference field.
Figure~\ref{fig:sr64_fields} compares the
reconstructions for an unseen turbulent realization.

At this resolution, the coarse observation retains
primarily the large-scale variations of the vorticity
field, while intermediate- and small-scale structures
are largely unresolved. Consequently, bicubic
interpolation produces a strongly smoothed
reconstruction that fails to reproduce the
filamentary organization of the DNS reference.
In contrast, WDSR recovers numerous elongated
vorticity filaments, localized extrema, and
coherent structures, resulting in substantially
improved spatial agreement.

Quantitatively, the relative $L_2$ error decreases
from $75.69\%$ for bicubic interpolation to
$14.30\%$ for WDSR, corresponding to an
$81.1\%$ reduction. The root-mean-square error
decreases from $10.4891$ to $1.9824$, while
the mean absolute error decreases from
$8.3182$ to $1.0062$. Similarly, the peak
signal-to-noise ratio increases from
$20.48$ to $34.95$~dB. The lower row of
Fig.~\ref{fig:sr64_fields} further demonstrates
that WDSR substantially reduces the
$Q$-criterion errors associated with
coherent vortical structures and regions
of strong velocity gradients.

Although the WDSR relative $L_2$ error increases
from $4.25\%$ at $\times32$ to $14.30\%$
at $\times64$, the reconstruction remains
considerably more accurate than bicubic
interpolation. This deterioration reflects
the increasing ambiguity of recovering
fine-scale structures from progressively
coarser observations.

\begin{figure}[ht!]
    \centering
    \includegraphics[width=0.75\linewidth]{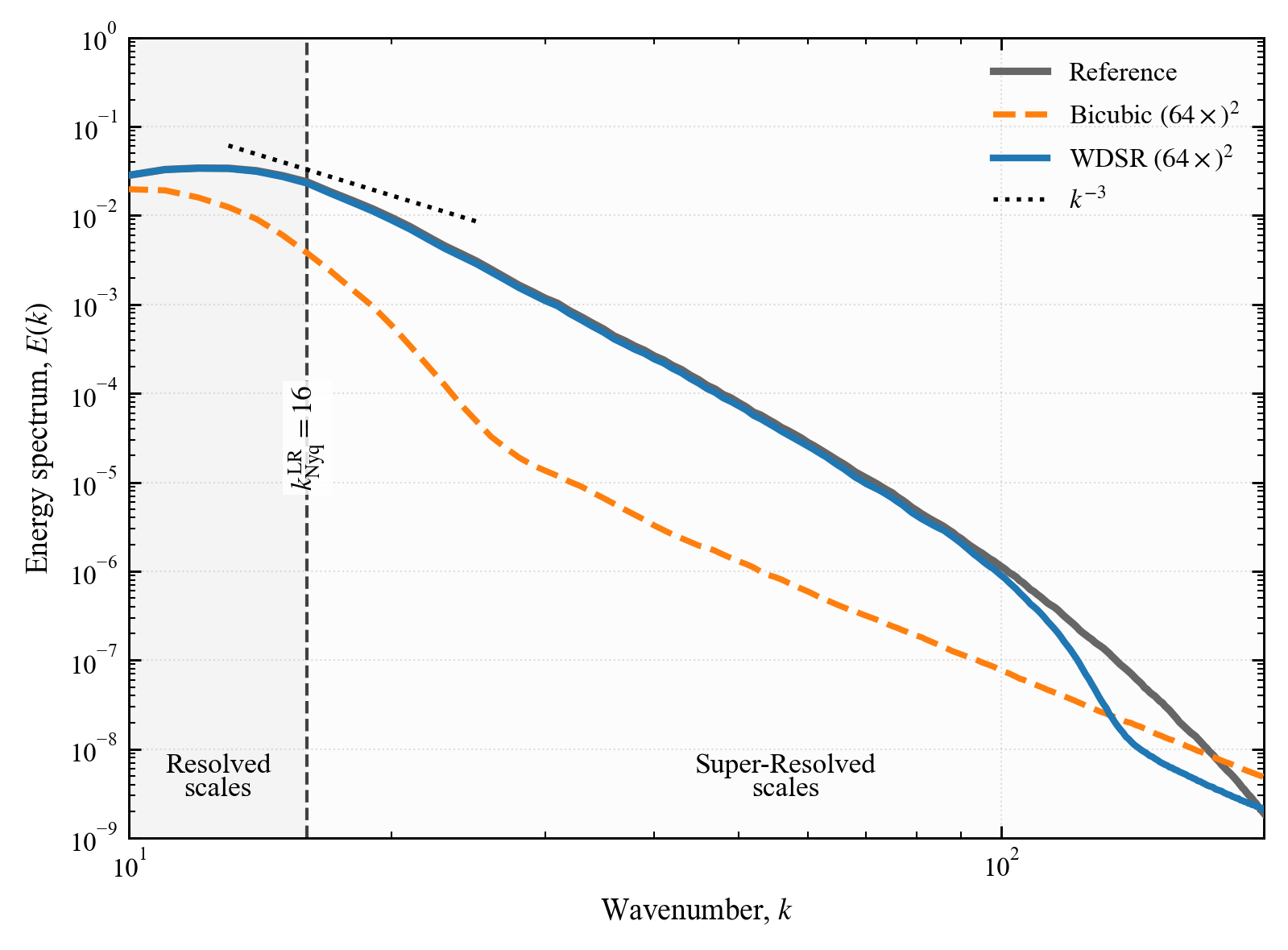}
    \caption{
    Angle-averaged energy spectrum for the
    $\times64$ reconstruction of an unseen
    turbulent realization at $Re=1000$.
    The DNS reference is compared with
    bicubic interpolation and the
    single-realization WDSR reconstruction.
    The vertical dashed line indicates
    the low-resolution Nyquist wavenumber,
    $k_{\mathrm{Nyq}}^{\mathrm{LR}}=16$,
    beyond which spatial scales are not
    directly represented in the coarse
    observation. The $k^{-3}$ line
    denotes a reference scaling.
    }
    \label{fig:spectrum64}
\end{figure}

The corresponding angle-averaged energy spectra
are presented in Fig.~\ref{fig:spectrum64}.
For the $32\times32$ observation, the
low-resolution Nyquist wavenumber is
$k_{\mathrm{Nyq}}^{\mathrm{LR}}=16$.

The WDSR spectrum follows the DNS reference
over a substantial wavenumber interval extending
beyond $k_{\mathrm{Nyq}}^{\mathrm{LR}}$,
indicating improved recovery of the energy
distribution at unresolved scales. In contrast,
bicubic interpolation exhibits pronounced
spectral attenuation, particularly at
higher wavenumbers.

Nevertheless, the WDSR spectrum eventually
departs from the DNS reference and decays
more rapidly at sufficiently high wavenumbers.
This departure is more pronounced than
in the $\times32$ case, indicating a
reduced range of accurately reconstructed
scales under more severe coarsening.
Importantly, agreement in the energy
spectrum establishes spectral consistency
but does not, by itself, demonstrate
realization-specific recovery of individual
Fourier modes.

\begin{figure}[ht!]
    \centering
    \includegraphics[width=0.75\linewidth]{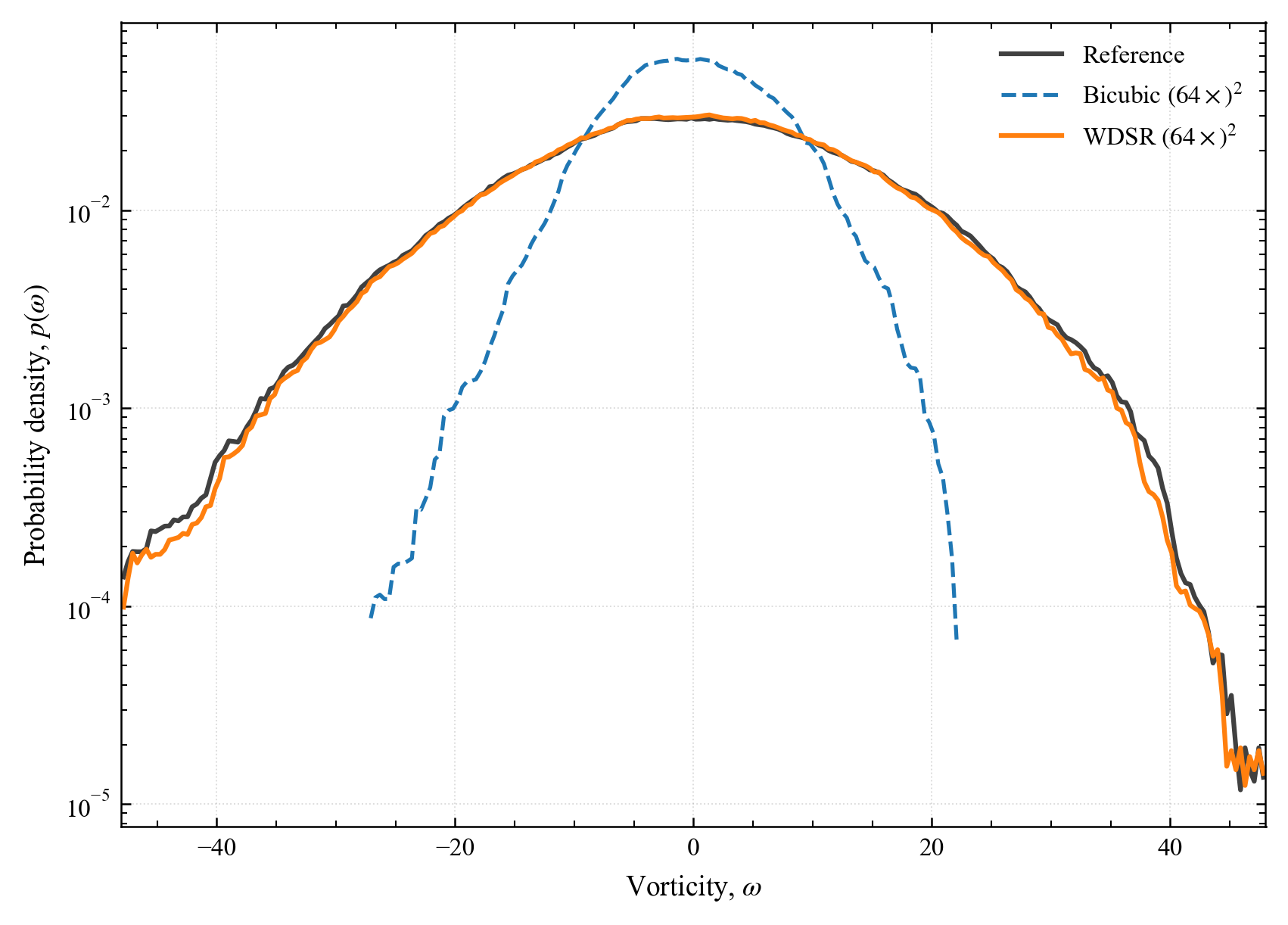}
    \caption{
    Probability density function of vorticity
    for the $\times64$ reconstruction at
    $Re=1000$, comparing the DNS reference,
    bicubic interpolation, and
    single-realization WDSR reconstruction.
    }
    \label{fig:pdf64}
\end{figure}

The one-point vorticity statistics are
examined in Fig.~\ref{fig:pdf64}.

The WDSR prediction reproduces the DNS
vorticity distribution over a broad range
of amplitudes, including the low-probability
tails associated with intense fluctuations.
Bicubic interpolation, by comparison,
produces a narrower distribution and
underrepresents large-amplitude events.
The improved agreement of the WDSR PDF
therefore indicates that the reconstructed
field preserves substantially more of the
one-point vorticity statistics despite
the severe reduction in input resolution.

Overall, the $\times64$ results demonstrate
that single-realization learning can provide
substantial improvements over conventional
interpolation in physical-space accuracy,
spectral energy distribution, and vorticity
statistics. However, the increased
reconstruction error and earlier
high-wavenumber spectral departure,
relative to $\times32$, reveal the
limitations imposed by extreme
spatial under-resolution.

\subsection{Scale-dependent Fourier coherence}
\label{subsec:fourier_coherence}

Agreement between reconstructed and reference energy spectra
does not necessarily imply recovery of the realization-specific
organization of unresolved turbulent structures. Two fields
may exhibit similar spectral energy distributions while
differing substantially in the phases and relative
organization of their Fourier modes. To distinguish spectral
fidelity from realization-specific correspondence, we
evaluate the scale-dependent Fourier coherence between
the reconstructed and reference vorticity fields.

Let $\widehat{\omega}_{\mathrm{ref}}(\boldsymbol{k})$ and
$\widehat{\omega}_{\mathrm{rec}}(\boldsymbol{k})$ denote the
two-dimensional Fourier coefficients of the reference and
reconstructed vorticity fields, respectively, where
$\boldsymbol{k}=(k_x,k_y)$ is the wavenumber vector.
The shell-averaged normalized Fourier cross-correlation
is defined as
\begin{equation}
\gamma(k)=
\frac{
\left|
\left\langle
\widehat{\omega}_{\mathrm{ref}}(\boldsymbol{k})
\widehat{\omega}_{\mathrm{rec}}^{*}(\boldsymbol{k})
\right\rangle_k
\right|
}{
\sqrt{
\left\langle
|\widehat{\omega}_{\mathrm{ref}}(\boldsymbol{k})|^2
\right\rangle_k
\left\langle
|\widehat{\omega}_{\mathrm{rec}}(\boldsymbol{k})|^2
\right\rangle_k
}
},
\label{eq:fourier_coherence}
\end{equation}
where $(\cdot)^*$ denotes complex conjugation and
$\langle\cdot\rangle_k$ indicates averaging over Fourier
modes within the same radial wavenumber shell.
The coherence satisfies $0\leq\gamma(k)\leq1$,
with values approaching unity indicating strong
normalized cross-spectral correspondence.
Unlike the energy spectrum, this diagnostic incorporates
the relative organization of complex Fourier coefficients,
although it does not independently establish agreement
in spectral amplitudes.

\begin{figure}[t]
    \centering
    \includegraphics[
        width=0.75\linewidth
    ]{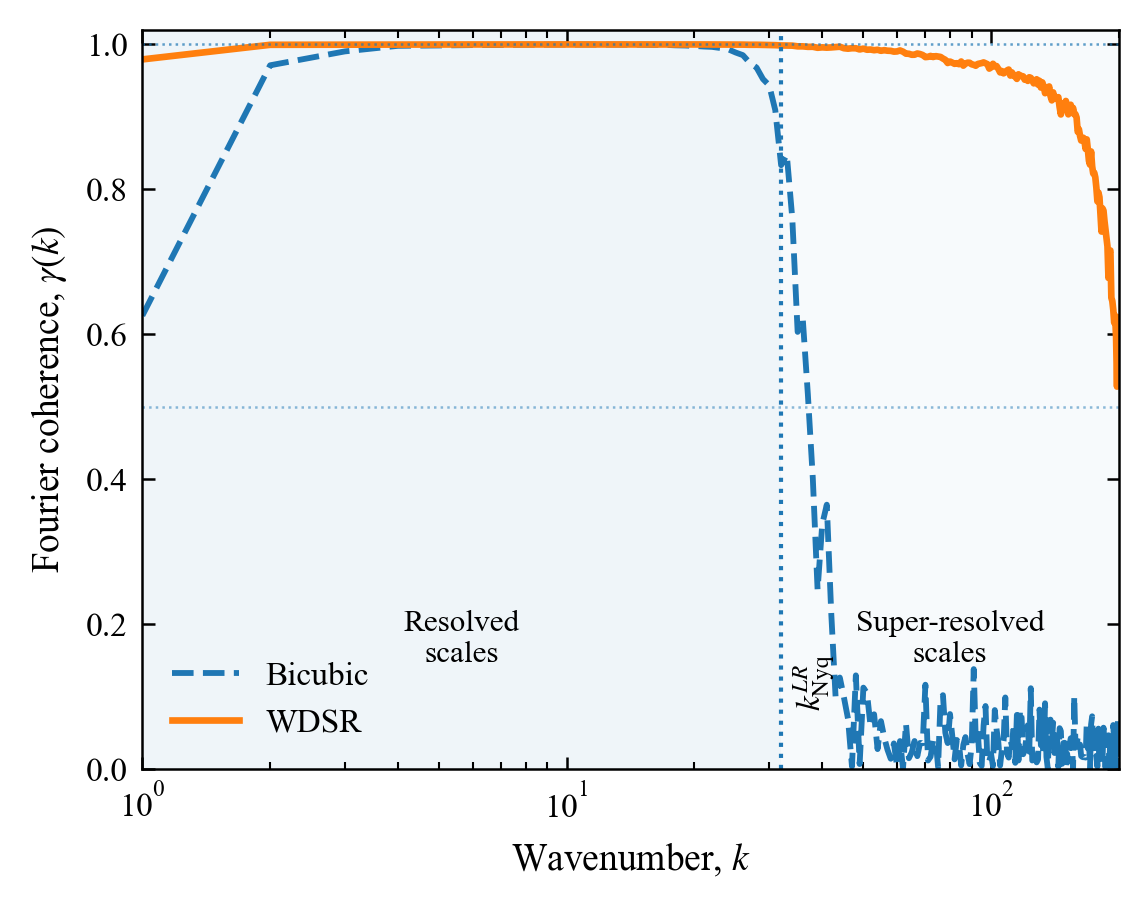}
    \caption{
    Scale-dependent Fourier coherence between the reconstructed
    and reference vorticity fields for $\times32$ super-resolution.
    }
    \label{fig:fourier_coherence_x32}
\end{figure}

Figure~\ref{fig:fourier_coherence_x32} presents the
Fourier coherence for $\times32$ super-resolution.
Bicubic interpolation exhibits high coherence at low
wavenumbers but deteriorates rapidly near and beyond
the low-resolution Nyquist limit,
$k_{\mathrm{Nyq}}^{\mathrm{LR}}=32$.
In contrast, WDSR maintains near-unity coherence
throughout much of the resolved range and retains
substantial cross-spectral correspondence beyond
the low-resolution cutoff. This indicates that the
learned reconstruction recovers not only high-wavenumber
spectral content but also Fourier information correlated
with the particular DNS realization.


\begin{figure}[t]
    \centering
    \includegraphics[
        width=0.75\linewidth
    ]{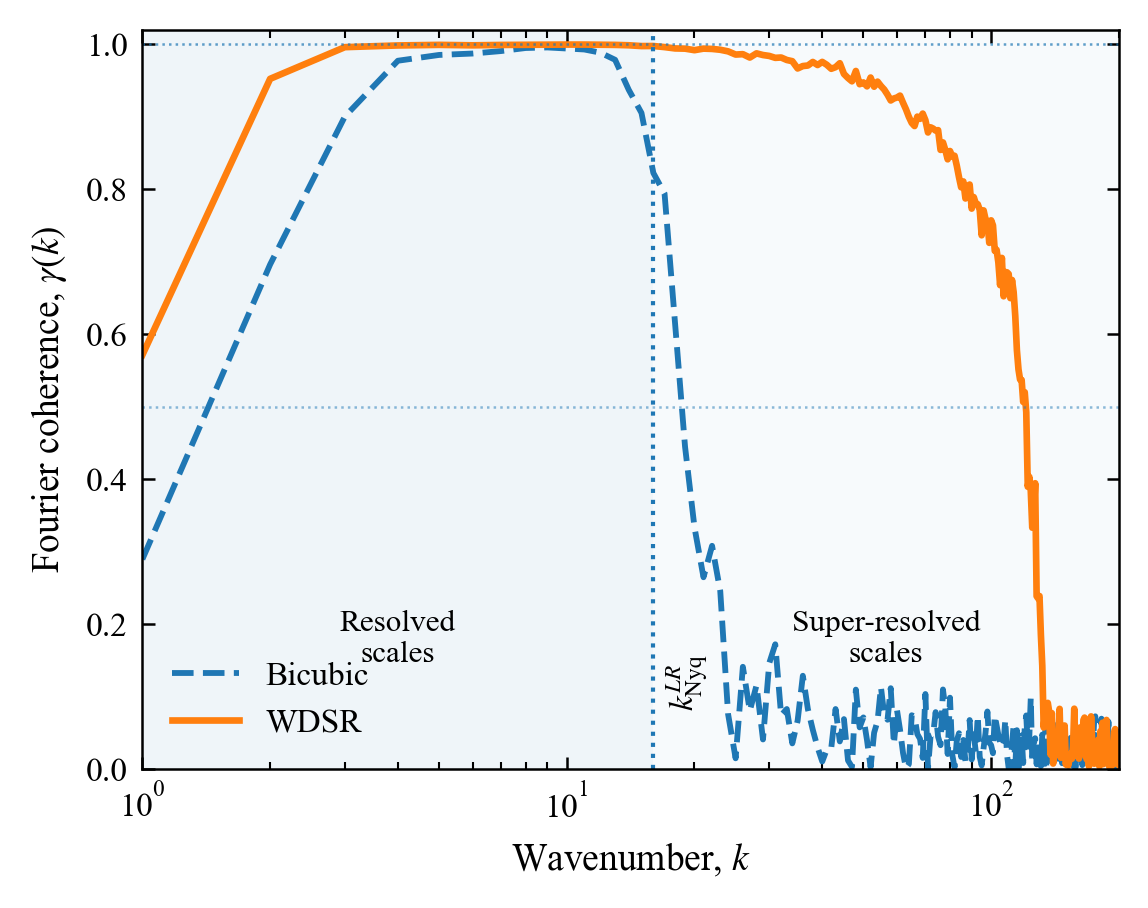}
    \caption{
    Scale-dependent Fourier coherence between the reconstructed
    and reference vorticity fields for $\times64$ super-resolution.
    }
    \label{fig:fourier_coherence_x64}
\end{figure}

The corresponding $\times64$ results are shown in
Fig.~\ref{fig:fourier_coherence_x64}.
With the low-resolution Nyquist wavenumber reduced
to $k_{\mathrm{Nyq}}^{\mathrm{LR}}=16$, a larger
portion of the turbulent spectrum must be inferred
from the coarse observation. WDSR nevertheless
maintains substantial coherence beyond this cutoff,
whereas bicubic interpolation exhibits a pronounced
reduction in cross-spectral correlation.
Compared with $\times32$, however, the WDSR
coherence decreases at lower absolute wavenumbers
and deteriorates more rapidly toward the finest
scales, indicating a narrower range of
realization-specific spectral recovery under
more severe spatial coarsening.


The persistence of coherence beyond the
low-resolution Nyquist limit in both cases
provides evidence that the learned mapping
captures coarse-to-fine relationships that
generalize to unseen turbulent realizations.
Nevertheless, the progressive loss of coherence
at higher wavenumbers reveals the limitations
of deterministic reconstruction under severe
spatial under-resolution. Moreover, because
Fourier coherence is normalized, strong
cross-spectral correspondence does not
necessarily imply accurate spectral amplitudes
or pointwise reconstruction. The coherence
results must therefore be interpreted alongside
the energy spectra, vorticity statistics, and
physical-space errors presented earlier.
Together, these complementary diagnostics
provide a more complete assessment of both
the statistical fidelity and realization-specific
accuracy of extreme-scale turbulent
super-resolution.

\subsection{Single-realization versus multi-realization learning}
\label{sec:single_vs_multi}

The preceding results establish that the proposed framework can perform
extreme super-resolution using training information extracted from only
one high-resolution turbulent realization. A central question,
however, is whether this strategy provides a meaningful advantage over
conventional WDSR training in which additional independent
high-resolution realizations are supplied to the model.

To investigate this question, we perform a data-scaling experiment for
the $\times32$ reconstruction problem. Separate WDSR models are trained
conventionally using $1$, $25$, $50$, and $100$ complete
high-resolution turbulent snapshots. These models do not employ the
overlapping patch-extraction and overlap-add reconstruction strategy of
the proposed framework. They are compared against the WDSR model
trained using overlapping patches extracted from a single
high-resolution realization. All methods are evaluated using the same
unseen turbulent field.

\begin{figure}[ht!]
    \centering
    \includegraphics[width=0.75\linewidth]{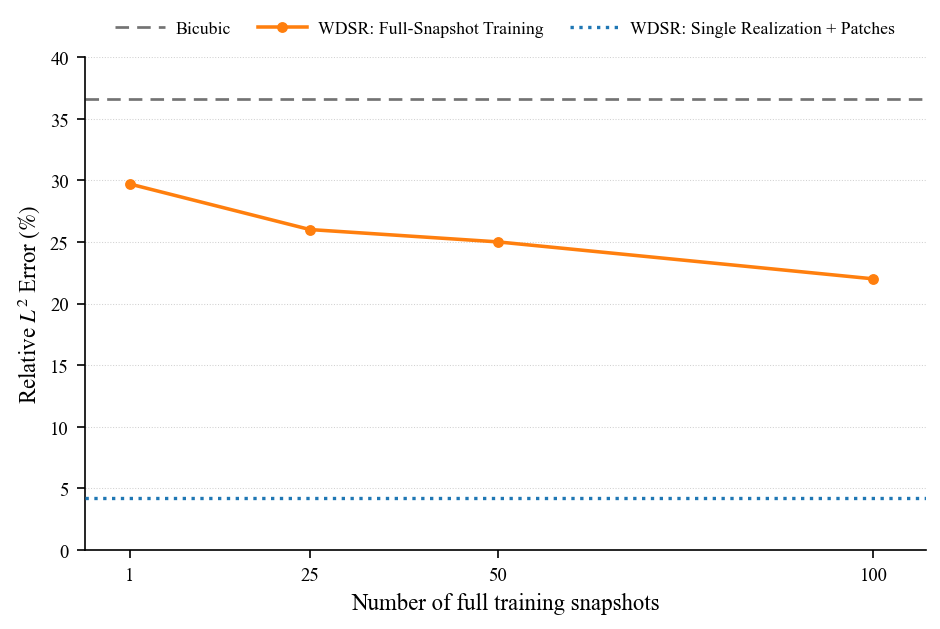}
    \caption{
    Influence of training-data availability and learning strategy on
    $\times32$ turbulent super-resolution.
    }
    \label{fig:data_scaling}
\end{figure}

Figure~\ref{fig:data_scaling} summarizes the resulting relative $L_2$
reconstruction errors.

Bicubic interpolation yields a relative $L_2$ error of approximately
$36.63\%$. Conventional WDSR training using a single complete snapshot
reduces the error to approximately $29.7\%$, demonstrating that the
learned reconstruction itself provides an improvement over interpolation.
Increasing the conventional training dataset to $25$, $50$, and $100$
complete snapshots further reduces the error to approximately $26\%$,
$25\%$, and $22\%$, respectively. The monotonic reduction in error
demonstrates the expected benefit of providing conventional WDSR with a
larger ensemble of independent high-resolution flow states.

The improvement obtained through conventional dataset scaling,
however, is considerably smaller than that obtained by exploiting the
internal spatial information of a single realization. The proposed
overlapping-patch WDSR framework achieves a relative $L_2$ error of
approximately $4.25\%$ using only one high-resolution training
snapshot. Compared with conventional WDSR trained using one complete
snapshot, the error is reduced by approximately $86\%$. More
strikingly, the conventional WDSR model trained using $100$
high-resolution realizations still exhibits an error of approximately
$22\%$, more than five times the error obtained using the proposed
single-realization strategy.

These results demonstrate that the number of global turbulent
realizations alone is not an adequate measure of the amount of useful
information presented to the learning algorithm. A single
high-resolution realization contains a large number of spatially
distributed local configurations. By exploiting statistical
homogeneity through overlapping patch extraction, the proposed
framework exposes the network to many localized coarse-to-fine
relationships contained within that realization. For the present
$\times32$ reconstruction problem, this internal spatial sampling is
substantially more effective than increasing the conventional
full-field training dataset from one to one hundred independent
snapshots.

The comparison in Fig.~\ref{fig:data_scaling} must nevertheless be
interpreted carefully. It is intentionally a comparison between two
different learning strategies rather than a controlled ablation of
snapshot count alone. The proposed model and the conventional
full-snapshot models differ both in the amount of high-resolution data
used and in the manner in which that information is presented to the
network. Consequently, the observed improvement should be attributed to
the complete single-realization patch-learning framework rather than
solely to the reduction in the number of training snapshots. This
distinction is important because the objective of the present study is
precisely to determine whether intensive exploitation of spatial
information within one realization can provide an effective alternative
to conventional dataset scaling.

\subsection{Effect of overlap-add reconstruction}
\label{sec:overlap_ablation}

To assess the influence of patch assembly on reconstruction
accuracy, we compare three inference strategies:
(i) non-overlapping patch reconstruction,
(ii) $50\%$ overlapping reconstruction with uniform averaging,
and (iii) $50\%$ overlapping reconstruction with
Hann-window-weighted assembly. The latter follows the
normalized overlap-add formulation introduced in
Section~\ref{sec:patch_reconstruction}. All three strategies
employ the same trained WDSR model, allowing the influence
of the assembly procedure to be examined independently
of the learned reconstruction mapping.

\begin{figure*}[ht!]
    \centering
    \includegraphics[width=\textwidth]{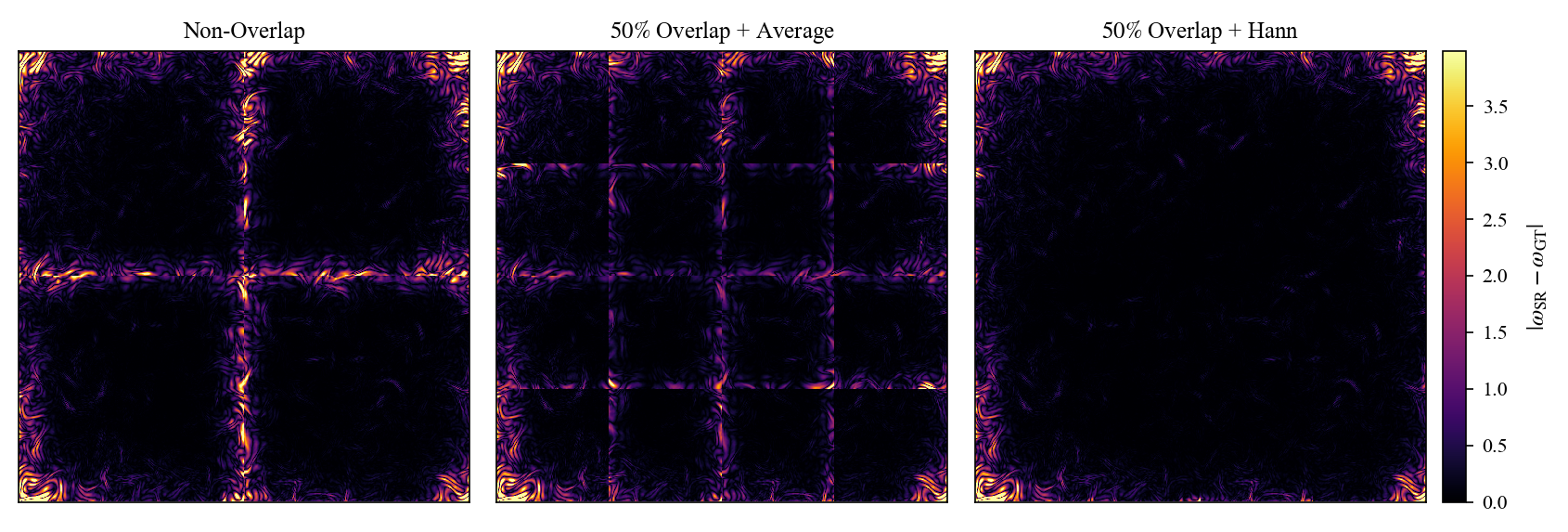}
    \caption{
    Effect of patch-assembly strategy on the spatial
    distribution of absolute vorticity reconstruction error,
    $|\omega^{\mathrm{SR}}-\omega^{\mathrm{GT}}|$.
    Results are shown for non-overlapping reconstruction
    (left), $50\%$ overlap with uniform averaging (center),
    and $50\%$ overlap with Hann-window weighting (right).
    }
    \label{fig:overlap_ablation}
\end{figure*}

Figure~\ref{fig:overlap_ablation} compares the spatial
distributions of absolute vorticity reconstruction error,
$|\omega^{\mathrm{SR}}-\omega^{\mathrm{GT}}|$, obtained using
the three assembly strategies. Non-overlapping reconstruction
exhibits pronounced grid-like error patterns concentrated
along patch interfaces, indicating reduced prediction
accuracy near patch boundaries. Introducing $50\%$ overlap
with uniform averaging mitigates these discontinuities,
although residual artifacts associated with the patch
geometry remain visible.

Hann-window-weighted assembly further suppresses these
boundary artifacts by preferentially weighting predictions
from patch interiors, where greater spatial context is
available. The resulting error distribution exhibits
substantially weaker signatures of the underlying patch
partitioning, indicating improved spatial consistency
across independently reconstructed regions.

These results demonstrate that the assembly strategy
influences the spatial distribution of reconstruction
errors, even when the underlying super-resolution model
remains unchanged. In particular, normalized Hann-window
overlap-add provides a more spatially consistent
full-field reconstruction than direct patch placement
or uniform overlap averaging. Accordingly, this strategy
is adopted throughout the present investigation.

\subsection{Robustness to noisy low-resolution observations}
\label{sec:noise_robustness}

The preceding results consider ideal low-resolution observations generated
directly from the high-resolution reference fields. In practical numerical
and experimental settings, however, coarse observations may additionally
contain measurement uncertainty, numerical contamination, or sensor noise.
We therefore examine the robustness of the learned reconstruction to
perturbations introduced directly into the low-resolution input.

Additive Gaussian noise is applied to the low-resolution field prior to
super-resolution, with noise levels ranging from $0\%$ to $20\%$. The trained
WDSR models are not retrained or fine-tuned for the noisy inputs. The noisy
fields are supplied directly to the reconstruction model, thereby providing
a stringent test of whether the multiscale mapping learned from the single
high-resolution realization remains effective when the resolved information
itself is perturbed.

\begin{figure}[ht!]
    \centering
    \includegraphics[width=0.75\linewidth]{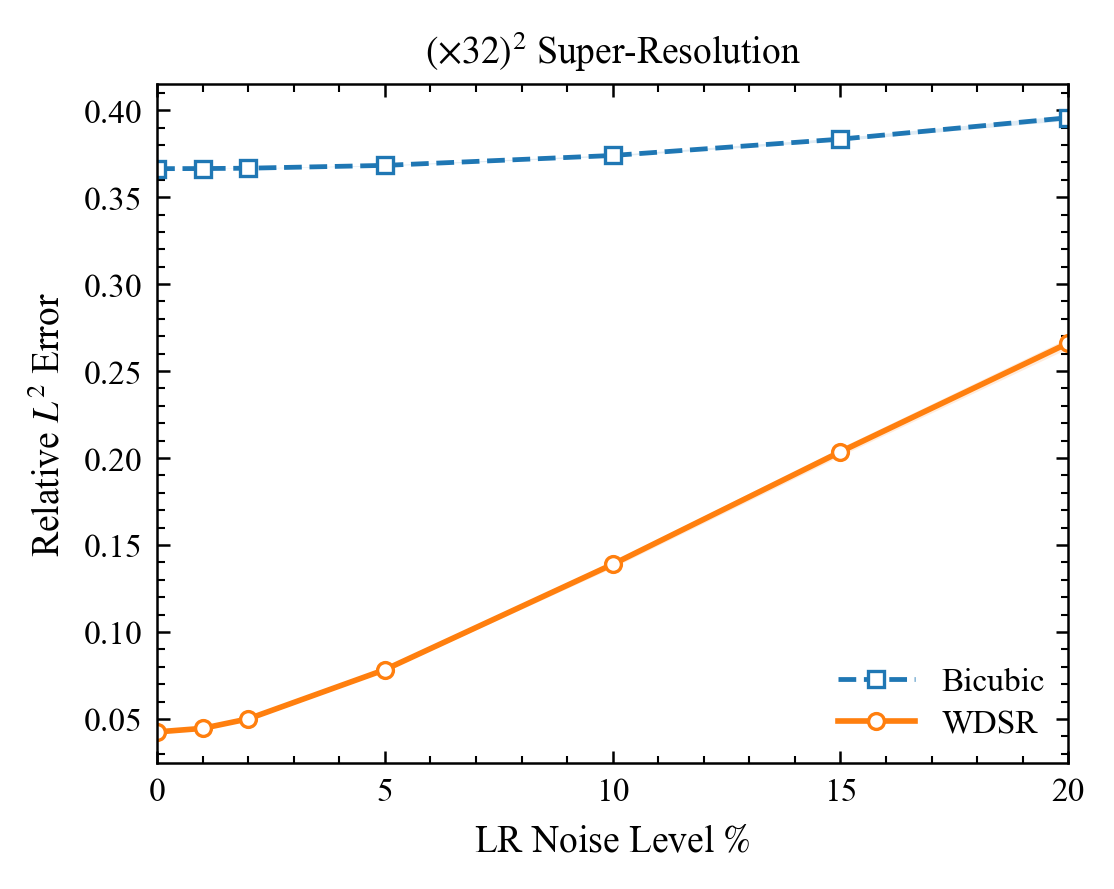}
    \caption{
    Sensitivity of the $\times32$ super-resolution reconstruction to noise
    in the low-resolution input.
    }
    \label{fig:noise32}
\end{figure}

Figure~\ref{fig:noise32} presents the resulting relative $L^2$ reconstruction
error for the $\times32$ case.

In the absence of input noise, the $\times32$ WDSR reconstruction exhibits
a relative $L^2$ error of approximately $4.25\%$, compared with approximately
$36.6\%$ for bicubic interpolation. At low noise levels, the WDSR error
increases only modestly, indicating that the learned reconstruction is not
excessively sensitive to small perturbations in the coarse observation.
As the noise amplitude is increased further, the error grows progressively,
reaching approximately $14\%$, $20\%$, and $27\%$ at noise levels of
$10\%$, $15\%$, and $20\%$, respectively.

Despite this degradation, WDSR remains substantially more accurate than
bicubic interpolation over the complete noise range. The bicubic error,
which is already large for the noise-free observation, changes only
moderately with increasing noise because interpolation is dominated by the
severe information loss associated with the original $\times32$ coarsening.
The learned reconstruction, by contrast, starts from a much lower error and
retains a significant advantage even when the low-resolution observation is
strongly contaminated.

\begin{figure}[ht!]
    \centering
    \includegraphics[width=0.75\linewidth]{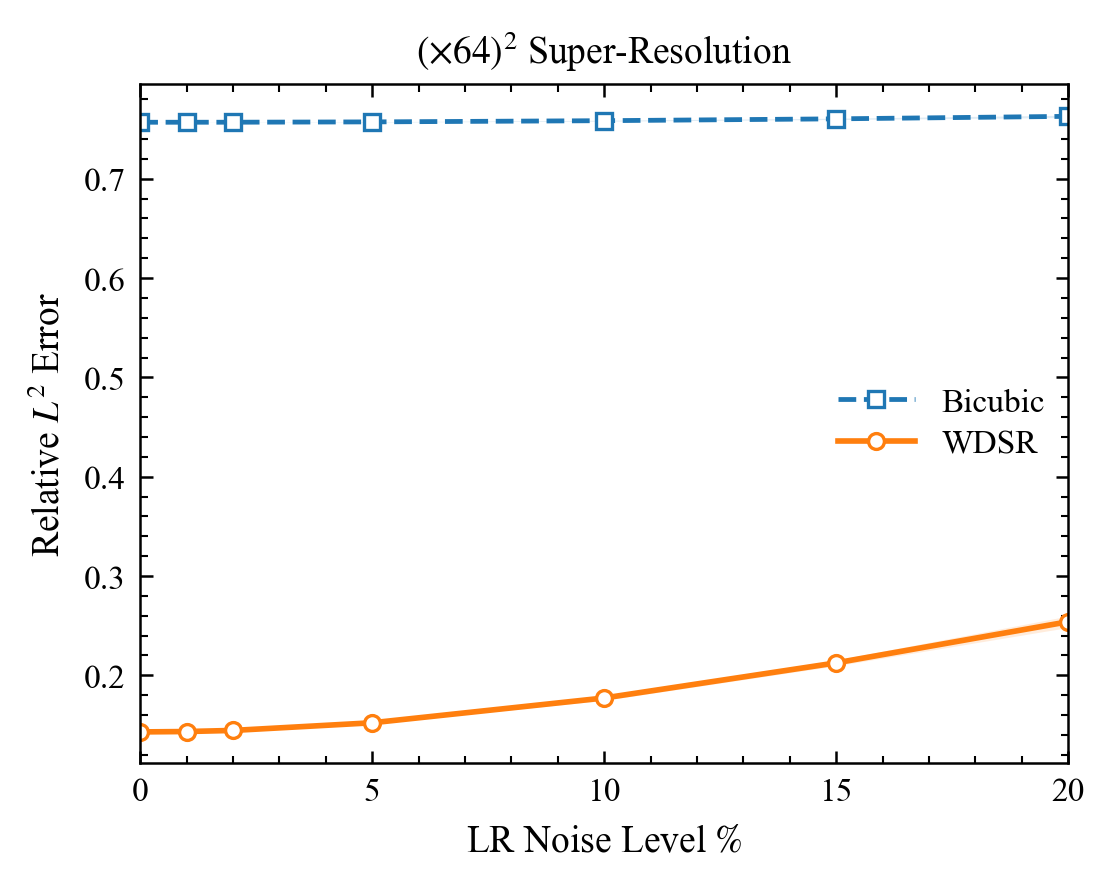}
    \caption{
    Sensitivity of the extreme $\times64$ super-resolution reconstruction
    to noise in the low-resolution input.
    }
    \label{fig:noise64}
\end{figure}

A similar analysis is performed for the more demanding $\times64$ problem,
as shown in Fig.~\ref{fig:noise64}.

The $\times64$ case begins from a larger reconstruction error because the
$32\times32$ observation retains substantially less information than the
$64\times64$ input employed for $\times32$ super-resolution. Nevertheless,
the WDSR reconstruction remains considerably more accurate than bicubic
interpolation. The error increases gradually from approximately $14\%$ for
the noise-free observation to approximately $18\%$, $21\%$, and $25\%$ for
noise levels of $10\%$, $15\%$, and $20\%$, respectively. Bicubic
interpolation, in comparison, remains near a relative $L^2$ error of
approximately $75\%$ over the same range.

Interestingly, the relative increase in WDSR error with input noise is less
pronounced for $\times64$ than for $\times32$. This behavior should not be
interpreted as indicating that the $\times64$ reconstruction is intrinsically
more accurate or more noise tolerant. Rather, at $\times64$, the dominant
source of reconstruction difficulty is already the extreme loss of spatial
information associated with reducing the reference field to only
$32\times32$ samples. Consequently, additional perturbations to the coarse
input produce a comparatively smaller increase relative to the existing
under-resolution error.

Taken together, Figs.~\ref{fig:noise32} and \ref{fig:noise64} demonstrate
that the single-realization WDSR framework does not rely on perfectly
noise-free coarse observations. Reconstruction accuracy decreases
continuously as the quality of the input deteriorates, as expected, but no
abrupt breakdown is observed over the range examined. More importantly, the
learned reconstruction maintains a substantial advantage over conventional
interpolation even under strong low-resolution perturbations. These results
provide additional evidence that the multiscale relationships extracted from
a single turbulent realization remain useful under imperfect observational
conditions.
\subsection{Physical interpretation and limits of extreme-scale reconstruction}
\label{sec:physical_interpretation}

The combined physical-space, spectral, and statistical results provide
a consistent picture of the reconstruction process. At $\times32$, the
$64\times64$ observation contains only $1/1024$ of the number of
spatial samples in the $2048\times2048$ reference field. Nevertheless,
the single-realization WDSR model reconstructs the dominant filamentary
vorticity organization, substantially reduces errors in derived
coherent-structure diagnostics, closely reproduces the reference
spectrum over a broad wavenumber range, and recovers the vorticity PDF,
including its low-probability tails.

The $\times64$ experiment represents a substantially stronger test of
the same hypothesis. In this case, only $1/4096$ of the reference
spatial samples are retained, and the low-resolution Nyquist wavenumber
is reduced to $k_{\mathrm{Nyq}}^{\mathrm{LR}}=16$. A large fraction of
the spectral range represented in the DNS field is therefore absent
from the observation. Despite this severe information deficit, the
WDSR model reconstructs coherent fine-scale organization, extends the
energy spectrum substantially beyond the low-resolution cutoff, and
closely reproduces the reference vorticity distribution.

The reconstruction should not, however, be interpreted as deterministic
recovery of all information removed during coarse graining. At extreme
scale separation, multiple fine-scale configurations may be compatible
with a given coarse observation. The WDSR model instead provides a
conditional reconstruction informed by two sources of information: the
resolved state supplied by the low-resolution observation and the
statistical relationships between coarse and fine scales learned from
the training realization. The increasing high-wavenumber discrepancy
observed for $\times64$ is consistent with this interpretation and
identifies a practical limit to the scales that can be inferred
reliably.

The results therefore support the central hypothesis of the present
study: for statistically homogeneous turbulence, a single
high-resolution realization can contain substantially more usable
learning information than its designation as a single snapshot might
suggest. Through overlapping spatial sampling, one realization provides
many localized manifestations of the underlying multiscale turbulent
organization. These internal statistics can be exploited to construct a
reconstruction prior that generalizes to an independent turbulent
state, even when the observation is severely under-resolved.

The comparison with conventional multi-realization training further
suggests that, under severe data scarcity, how the available
high-resolution information is exploited can be at least as important
as the number of independent realizations available. Rather than
relying exclusively on increasingly large high-fidelity datasets, the
present results demonstrate the potential of realization-centric
learning, in which the spatial information embedded within an available
turbulent field is systematically exploited to support extreme-scale
reconstruction.

\subsection{Reconstruction limitations at higher Reynolds number}
\label{sec:high_re_limitations}

To examine the applicability of single-realization
learning to more demanding turbulent regimes, we
consider an additional reconstruction experiment
at $Re=16000$. Compared with the $Re=1000$
configuration, the higher-Reynolds-number flow
exhibits a broader range of spatial scales and
more pronounced fine-scale vorticity structures,
providing a useful test of the reconstruction
capabilities of the proposed framework.

\begin{figure*}[ht!]
    \centering
    \includegraphics[width=\textwidth]{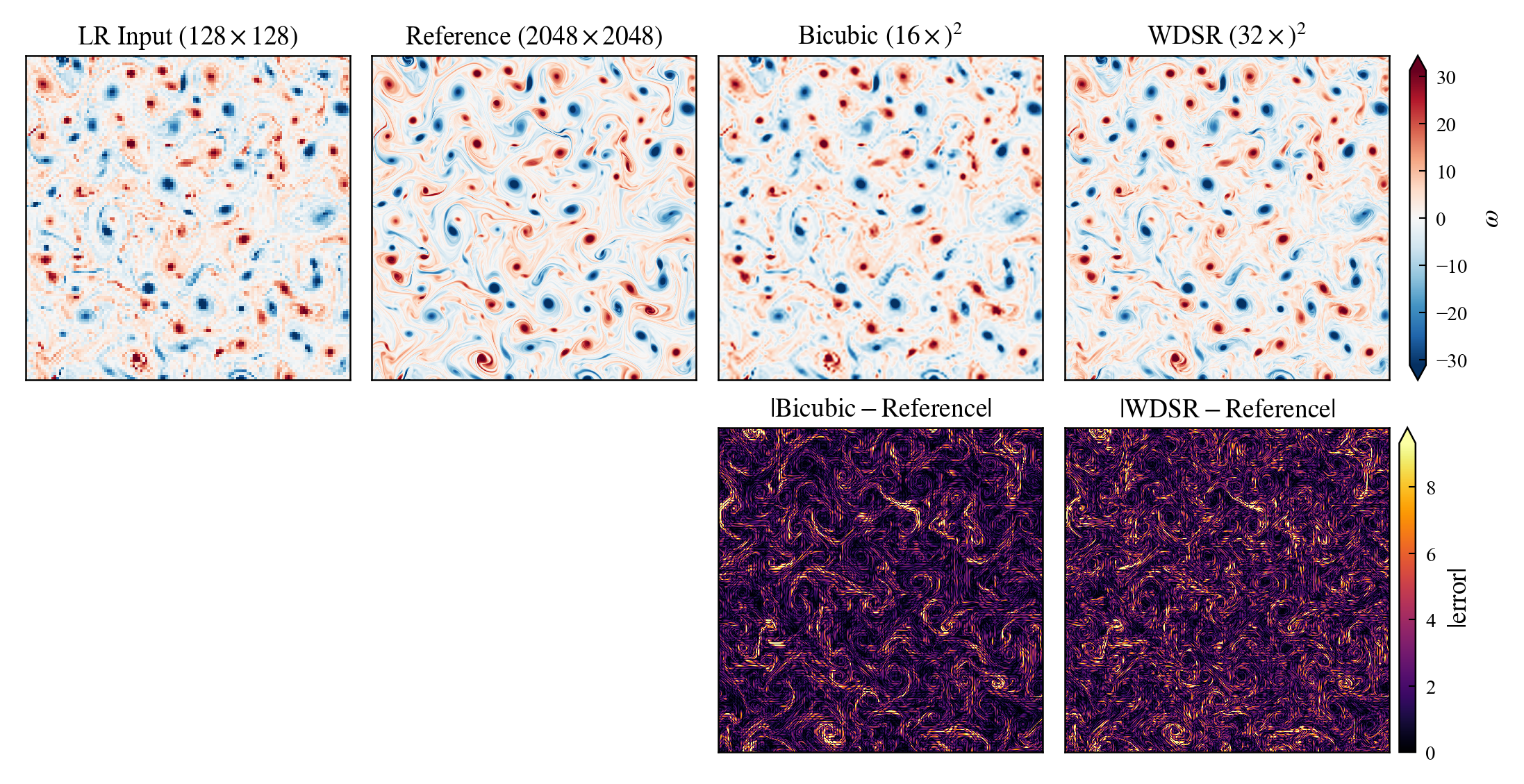}
    \caption{
    Super-resolution reconstruction of an unseen
    homogeneous isotropic turbulence realization
    at $Re=16000$ with a scale factor of
    $\times16$. The low-resolution observation,
    DNS reference, bicubic interpolation, and
    WDSR reconstruction are compared alongside
    the corresponding absolute reconstruction
    errors.
    }
    \label{fig:re16000_fields}
\end{figure*}

The high-resolution vorticity field is represented
on a $2048\times2048$ grid and reconstructed
from a $128\times128$ observation, corresponding
to a super-resolution factor of $\times16$.
Figure~\ref{fig:re16000_fields} compares the
low-resolution input, DNS reference, bicubic
interpolation, and WDSR reconstruction for
an unseen turbulent realization.

As shown in Fig.~\ref{fig:re16000_fields},
bicubic interpolation preserves the dominant
large-scale organization but substantially
attenuates the finer vorticity structures.
The WDSR reconstruction captures a greater
portion of the filamentary organization and
exhibits improved agreement with the DNS
reference. Nevertheless, localized discrepancies
remain near thin filaments and regions of
strong vorticity gradients, where small errors
in position or amplitude can produce
appreciable pointwise differences.

\begin{figure}[ht!]
    \centering
    \includegraphics[width=0.75\linewidth]{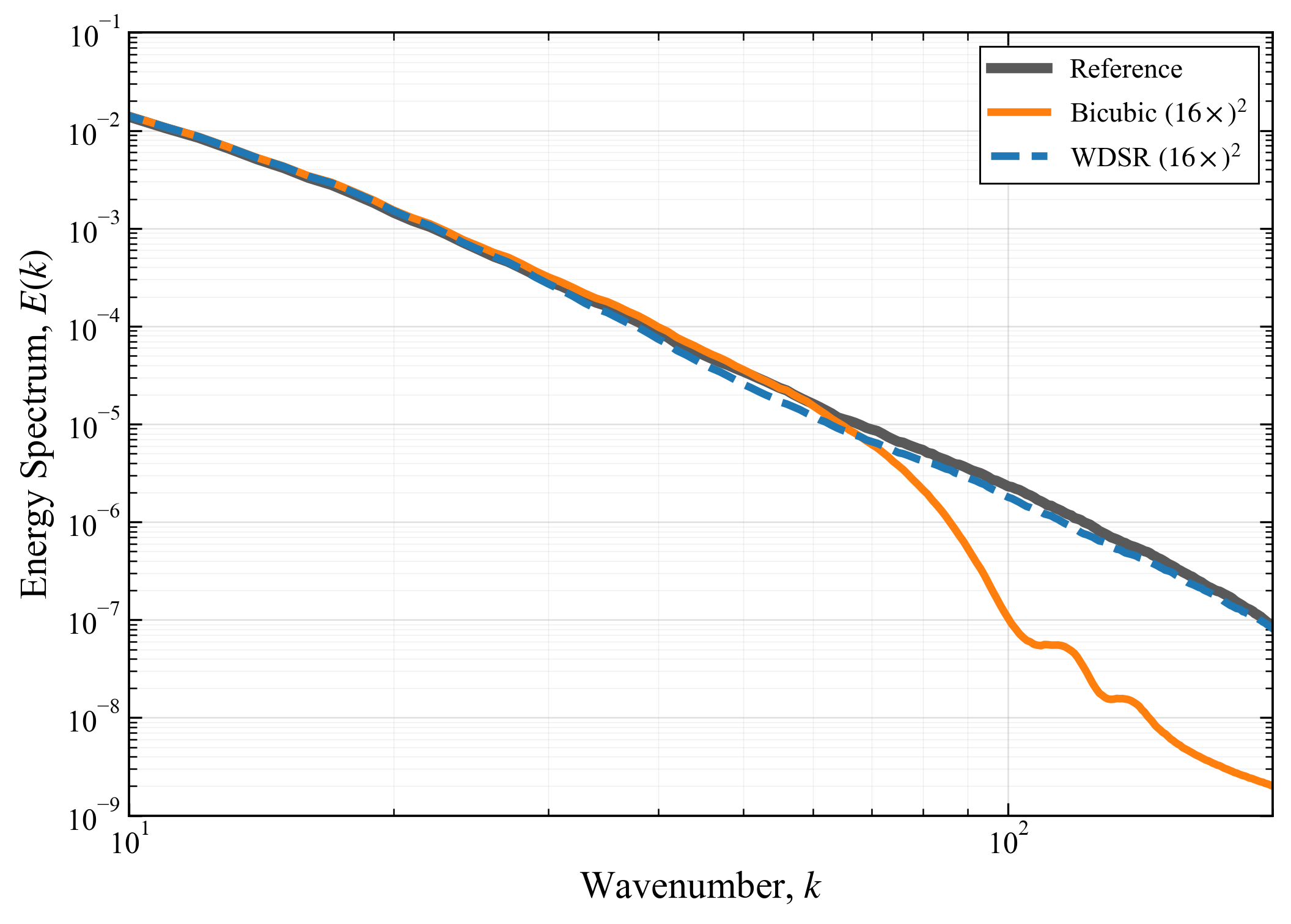}
    \caption{
    Angle-averaged energy spectrum for the
    $\times16$ super-resolution experiment
    at $Re=16000$, comparing the DNS
    reference, bicubic interpolation, and
    single-realization WDSR reconstruction.
    }
    \label{fig:re16000_spectrum}
\end{figure}

The corresponding angle-averaged energy
spectra are presented in
Fig.~\ref{fig:re16000_spectrum}.

The WDSR spectrum follows the DNS reference
over a substantially broader wavenumber range
than bicubic interpolation, which exhibits
pronounced attenuation at smaller spatial
scales. However, the WDSR spectrum also
departs from the reference at sufficiently
high wavenumbers, indicating incomplete
recovery of the finest resolved scales.
This behavior is consistent with the
localized reconstruction errors observed
in physical space, although spectral
agreement alone does not establish
realization-specific correspondence
between individual Fourier modes.

These results illustrate both the capabilities
and limitations of the present deterministic
framework. The network reconstructs
substantial fine-scale structure from
a severely under-resolved observation,
but its accuracy is reduced for features
whose spatial organization is weakly
constrained by the available coarse-scale
information. Because the Reynolds number
and super-resolution factor differ from
those considered in the preceding
experiments, this comparison should be
interpreted as an additional
high-Reynolds-number reconstruction test
rather than a controlled assessment
of Reynolds-number dependence.

\subsection{Discussion of reconstruction limitations}
\label{sec:re_dependence}

The reconstruction of unresolved turbulent
scales is fundamentally an underdetermined
inverse problem. Spatial coarsening removes
information about fine-scale fluctuations,
and distinct high-resolution realizations
may produce similar low-resolution
observations. Consequently, the relationship
between the coarse and fine fields is more
generally characterized by a conditional
distribution,

\begin{equation}
p\left(
    \omega^{HR}\mid\omega^{LR}
\right),
\label{eq:conditional_distribution}
\end{equation}

rather than a uniquely invertible mapping.
The deterministic WDSR model approximates
this relationship through a single
reconstruction conditioned on the
low-resolution observation.

The effectiveness of this approximation
depends on the information retained
by the coarse field and the diversity
of multiscale relationships represented
within the training realization.
Statistical homogeneity enables the
extraction of numerous localized training
samples from a single snapshot, but
does not guarantee that these samples
span all possible configurations of
the unresolved dynamics. In particular,
increasing the fraction of dynamically
relevant scales removed by coarse
graining can increase ambiguity in
the reconstruction of individual
fine-scale structures.

These considerations suggest several
possible extensions. Spectral and
gradient-sensitive loss functions could
place greater emphasis on small-scale
energy content and localized vorticity
gradients, while constraints derived
from the governing equations could
improve physical consistency.
Alternatively, probabilistic
reconstruction methods could represent
uncertainty in the unresolved scales
through multiple high-resolution
realizations conditioned on the
same coarse observation.

The present results therefore support
single-realization learning as a
data-efficient approach to turbulent
super-resolution while identifying
an important limitation of deterministic
reconstruction: accurate recovery of
unresolved statistics does not
necessarily imply exact recovery of
the fine-scale configuration of an
individual turbulent realization.

\section{Conclusion and Future Work}
\label{sec:conclusions}

This study investigated extreme-scale super-resolution
of two-dimensional homogeneous isotropic turbulence
using a single high-resolution realization for training.
The proposed framework combines localized patch-based
learning with a wide-activation deep super-resolution
(WDSR) network and Hann-window-weighted overlap-add
reconstruction to recover complete turbulent fields
from severely under-resolved observations. By exploiting
spatially recurring multiscale structures within a
statistically homogeneous flow, the framework learns
coarse-to-fine relationships without requiring
additional high-resolution training realizations.

The proposed approach achieved relative $L_2$ errors
of $4.25\%$ and $14.30\%$ for super-resolution
factors of $\times32$ and $\times64$, respectively,
compared with $36.64\%$ and $75.69\%$ for bicubic
interpolation. Beyond improved pointwise accuracy,
the reconstructed fields exhibited substantially
better agreement with the reference vortical
structures, turbulent energy spectra, and vorticity
probability distributions. Scale-dependent Fourier
coherence further demonstrated that the reconstructed
fields retain substantial realization-specific
cross-spectral correspondence beyond the
low-resolution Nyquist cutoff, indicating that
the learned mapping recovers information correlated
with the unresolved organization of previously
unseen turbulent realizations.

A comparison with conventional full-field WDSR
training further highlighted the effectiveness
of exploiting internal spatial information.
For the $\times32$ problem, the proposed
single-realization framework achieved a relative
$L_2$ error of $4.25\%$, whereas conventional
WDSR trained on $100$ complete high-resolution
snapshots yielded an error of approximately
$22\%$. Although these approaches differ in
training-data presentation and reconstruction
procedure, the comparison demonstrates the
practical advantage of the complete patch-based
learning framework under the investigated
conditions.

Despite these improvements, reconstruction accuracy
decreased with increasing spatial under-resolution,
with progressively larger discrepancies at the
finest scales. The additional high-Reynolds-number
experiment further illustrated the limitations
of deterministic reconstruction when unresolved
structures are insufficiently constrained by the
coarse observations. These findings emphasize
that agreement in spectral and statistical
properties does not necessarily imply exact
recovery of realization-specific fine-scale
structures.

Future work will investigate physics-informed
and spectrally constrained learning objectives,
uncertainty-aware reconstruction methods, and
extensions to three-dimensional and more complex
turbulent flows. The present findings demonstrate
that localized learning from a single turbulent
realization provides a promising data-efficient
strategy for full-domain extreme-scale
super-resolution.

\section*{Acknowledgments}
This work was supported in part by the AFOSR Grant FA9550-24-1-0327.

\section*{Data Availability}

The datasets and code used to generate the results presented
in this study are publicly available and can be accessed through
the following GitHub repository: \href{https://github.com/as26101999/Extreme-Scale-Super-Resolution-Using-a-Single-Realization}{Extreme-Scale-Super-Resolution}.

\bibliographystyle{unsrt}
\bibliography{references}

\end{document}